\documentclass[aps, prx, twocolumn, superscriptaddress, longbibliography, floatfix]{revtex4-2}
\usepackage{natbib}
\usepackage{amsmath} 
\usepackage{amssymb} 
\usepackage{graphicx} 
\usepackage{times}
\usepackage{color}
\usepackage{wasysym}
\usepackage{booktabs}
\usepackage{siunitx}

\begin{document}

\title{Electronic excitations in $5d^4$ $J$\,=\,0 Os$^{4+}$ halides studied by RIXS and 
	optical spectroscopy}

\author{P. Warzanowski}
\author{M. Magnaterra}
\author{P. Stein}
\author{G. Schlicht}
\affiliation{Institute of Physics II, University of Cologne, 50937 Cologne, Germany}
\author{Q. Faure}
\affiliation{ESRF, The European Synchrotron, 71 Avenue des Martyrs, CS40220, 38043 Grenoble Cedex 9, France}
\affiliation{Laboratoire L\'{e}on Brillouin, CEA, CNRS, Universit\'{e} Paris-Saclay, CEA-Saclay, 
	91191 Gif-sur-Yvette, France}
\author{Ch. J. Sahle}
\affiliation{ESRF, The European Synchrotron, 71 Avenue des Martyrs, CS40220, 38043 Grenoble Cedex 9, France}
\author{T.~Lorenz}
\affiliation{Institute of Physics II, University of Cologne, 50937 Cologne, Germany}
\author{P.~Becker}
\author{L.~Bohat\'{y}}
\affiliation{Sect. Crystallography, Institute of Geology and Mineralogy, University of Cologne, 50674 Cologne, Germany}
\author{M. Moretti Sala}
\affiliation{Dipartimento di Fisica, Politecnico di Milano, I-20133 Milano, Italy}
\author{G. Monaco}
\affiliation{Dipartimento di Fisica e Astronomia "Galileo Galilei", Universit\`{a} di Padova, I-35121 Padova, Italy}
\author{P.H.M. van Loosdrecht}
\author{M. Gr\"{u}ninger}
\affiliation{Institute of Physics II, University of Cologne, 50937 Cologne, Germany}

\begin{abstract}
We demonstrate that the cubic antifluorite-type halides K$_{\rm 2}$OsCl$_{\rm 6}$, 
K$_{\rm 2}$OsBr$_{\rm 6}$, and Rb$_{\rm 2}$OsBr$_{\rm 6}$ are excellent realizations 
of non-magnetic $J$\,=\,0 compounds. The magnetic susceptibility shows the corresponding 
Van-Vleck type behavior and no sign of defects. 
We investigate the electronic excitations with two complementary techniques, 
resonant inelastic x-ray scattering (RIXS) and optical spectroscopy. 
This powerful combination allows us to thoroughly study, e.g., on-site 
intra-$t_{2g}$ excitations and $t_{2g}$-to-$e_g$ excitations as well as 
inter-site excitations across the Mott gap and an exciton below the gap. 
In this way, we determine the electronic parameters with high accuracy, altogether yielding 
a comprehensive picture.
In K$_{\rm 2}$OsCl$_{\rm 6}$, we find the spin-orbit coupling constant $\zeta$\,=\,0.34\,eV, 
Hund's coupling $J_H$\,=\,0.43\,eV,  the onset of excitations across the Mott gap 
at $\Delta$\,=\,2.2\,eV, the cubic crystal-field splitting 10\,$Dq$\,=\,3.3\,eV, 
and the charge-transfer energy $\Delta_{\rm CT}$\,=\,4.6\,eV.\@ 
With $J_H/\zeta$\,=\,1.3, K$_{\rm 2}$OsCl$_{\rm 6}$ is in the intermediate-coupling regime. 
In a $t_{2g}$-only Kanamori picture, the above values correspond to 
$\zeta^{\rm eff}$\,=\,0.41\,eV and $J_H^{\rm eff}$\,=\,0.28\,eV, which is very close to 
results reported for related  $5d^4$ iridates.
In the tetragonal phase at 5\,K, the non-cubic crystal field causes a peak splitting 
of the $J$\,=\,1 state as small as 4\,meV.\@ 
Compared to K$_{\rm 2}$OsCl$_{\rm 6}$, the bromides K$_{\rm 2}$OsBr$_{\rm 6}$ and 
Rb$_{\rm 2}$OsBr$_{\rm 6}$ show about 12-14\,\% smaller values of 10\,$Dq$ and 
$\Delta_{\rm CT}$, while the spin-orbit-entangled intra-$t_{2g}$ excitations 
below 2\,eV and hence $\zeta$ and $J_H$ are reduced by less than 4\,\%. 
Furthermore, the Mott gap in K$_{\rm 2}$OsBr$_{\rm 6}$ is reduced to about 1.8\,eV.
\end{abstract}

\date{June 21, 2023}

\maketitle

The family of $5d$ transition-metal compounds features Mott-insulating quantum materials 
in which strong spin-orbit coupling plays the central role 
\cite{WitczakKrempa14,Rau16,Schaffer16,Takayama21,Khomskii21}. 
Prominent examples are $t_{2g}^5$ iridates with spin-orbit entangled $J$\,=\,$1/2$ moments 
\cite{Winter17,Cao18,Trebst22}. 
Compounds with edge-sharing IrO$_6$ octahedra have been predicted to show 
bond-directional Kitaev-type exchange couplings \cite{Jackeli09}. This has raised hopes 
to realize the Kitaev model on tricoordinated lattices, where strong exchange frustration
yields an intriguing quantum spin liquid \cite{Hermanns18,Takagi19,Motome20}. 
Experimentally, the dominant bond-directional character of exchange interactions has been 
demonstrated for honeycomb Na$_2$IrO$_3$ \cite{Chun15,Magnaterra23}. 
Remarkably, Kitaev exchange has been found to have a very different effect for $J$\,=\,1/2 
moments on an \textit{fcc} lattice with corner-sharing IrO$_6$ octahedra as realized 
in the double perovskite Ba$_2$CeIrO$_6$. There, antiferromagnetic Kitaev coupling 
counteracts the geometric frustration of isotropic Heisenberg exchange \cite{Revelli19fcc}.
For face-sharing IrO$_6$ octahedra as in the Ir$_2$O$_9$ dimer compounds 
Ba$_3$$M$Ir$_2$O$_9$, spin-orbit coupling competes with strong intra-dimer hopping 
that yields quasimolecular orbitals. 
Still, RIXS studies for $M$\,=\,Ce$^{4+}$ and In$^{3+}$ established the spin-orbit 
entangled $J_{\rm dim}$\,=\,0 and $J_{\rm dim}$\,=\,3/2 character of the 
respective quasimolecular magnetic moments \cite{Revelli19,Revelli22}.

For a $t_{2g}^4$ configuration in cubic symmetry, dominant spin-orbit coupling $\zeta$ is 
expected to yield a non-magnetic $J$\,=\,0 ground state.  
However, strong exchange interactions give rise to a dispersion of magnetic excited states, 
and if the dispersion is large enough, these may condense and drive magnetism of excitonic 
Van-Vleck-type that is also called 
singlet magnetism \cite{Khaliullin13,Akbari14,Meetei15,KhomskiiBook}. 
In this scenario, a magnetic amplitude mode equivalent to a Higgs mode is expected, which 
has been proposed for antiferromagnetic Ca$_2$RuO$_4$ \cite{Khaliullin13,Jain17}. 
In this layered $4d^4$ compound, one has to consider the tetragonal crystal-field splitting 
$\Delta_{\rm CF}$ \cite{Akbari14,Kunkemoeller15,Kunkemoeller17,Zhang17,Zhang20,Feldmaier20,Kim17}
and that spin-orbit coupling is smaller than in $5d$ materials such that the physics is 
governed by the ratio $\Delta_{\rm CF}/\zeta > 2$ \cite{Gretarsson19,Vergara22}. 
For dominant non-cubic $\Delta_{\rm CF}$, $d^4$ compounds turn into spin $S$\,=\,1 
magnets. 
Considering (nearly) cubic symmetry, the local intra-$t_{2g}$ excitations from the $J$\,=\,0 
state have been studied by RIXS in $4d^4$ K$_2$RuCl$_6$ \cite{Takahashi21} 
and in the $5d^4$ Ir$^{5+}$ double perovskites 
$A_2$$B$IrO$_6$ ($A$\,=\,Ba, Sr; $B$\,=\,Y, Gd, Lu, Sc) 
\cite{Yuan17,Kusch18,Nag18,Aczel22,Paramekanti18}.
Among these, Sr$_2$YIrO$_6$ and Ba$_2$YIrO$_6$ were reported to host magnetic order 
\cite{Cao14,Terzic17} which has been attributed to the presence of $5d^5$ Ir$^{4+}$ 
and $5d^3$ Ir$^{6+}$ defects \cite{Fuchs18}. 
Pyrochlore Yb$_2$Os$_2$O$_7$ with $5d^4$ Os$^{4+}$ ions exhibits a 
trigonal distortion and a defect-induced magnetic response \cite{Davies19}.
For K$_2$RuCl$_6$, the possible role of vibronic effects has been discussed \cite{Iwahara23}.

Here, we employ RIXS at the Os $L_3$ edge and optical spectroscopy to study 
stoichiometric single crystals of the $5d^4$ halides K$_2$OsCl$_6$ and $A_2$OsBr$_6$ 
with $A$\,=\,K and Rb. 
In the magnetic susceptibility, we do not find any indication of magnetic defects. 
Our results establish these materials as a reference for 
$J$\,=\,0 systems in the intermediate regime between $LS$ coupling for 
$J_H/\zeta \!\rightarrow \!\infty$ and $jj$ coupling for $J_H/\zeta \! \rightarrow\! 0$. 
We show that $J_H/\zeta$\,=\,1.3 is equivalent to $J_H^{\rm eff}/\zeta^{\rm eff}$\,=\,0.8 
in the Kanamori scheme \cite{Georges13}, which considers only $t_{2g}$ orbitals. 
Our result for $J_H^{\rm eff}/\zeta^{\rm eff}$ is in excellent agreement with the values 
reported for $5d^4$ iridates \cite{Yuan17,Kusch18,Nag18,Aczel22}.
The antifluorite-type Os compounds combine several properties which are advantageous for 
a precise determination of the electronic parameters. At room temperature, they show 
cubic symmetry \cite{Fergusson74,Armstrong78,Saura22}, and for the purpose of studying the 
local electronic excitations the compounds may be viewed as being composed of undistorted, 
well separated Os$X_6$ octahedra with, to first approximation, negligible interactions 
between them. 
This yields well-defined, narrow RIXS peaks even at high energy such as the charge-transfer 
excitations at 4.6, 5.7, and 8.0\,eV in K$_2$OsCl$_6$, allowing for a straightforward 
determination of the charge-transfer energy $\Delta_{\rm CT}$. 

\begin{figure}[t]
	\centering
	\includegraphics[width=0.98\columnwidth]{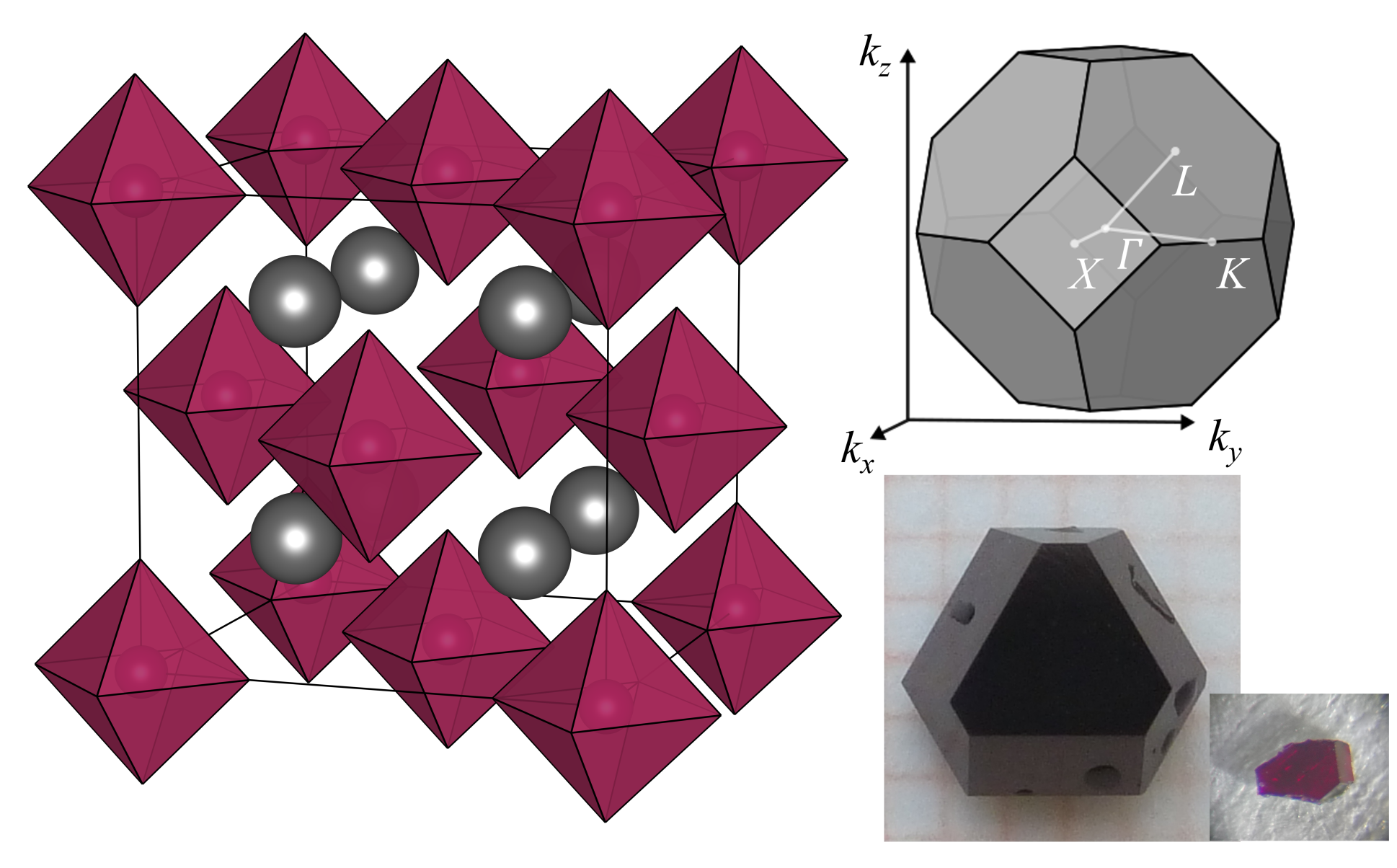}			
	\caption{{\bf Left: Face-centered cubic crystal structure of 
	\textit{A}$_2$Os\textit{X}$_6$ at 300\,K.}
	The Os$X_6$ octahedra ($X$\,=\,Cl, Br) are plotted in red and the $A^+$ ions 
	($A$\,=\,K, Rb) in gray. 
	Top right: sketch of the first Brillouin zone of the \textit{fcc} lattice with the 
	high-symmetry points $\Gamma$, $X$, $K$, and $L$. 
	Lower right: photos of a K$_2$OsBr$_6$ crystal and of the thin, transparent sample 
	of K$_2$OsCl$_6$ that has been studied in infrared transmittance.
	The size of the latter amounts to approx.\ 0.5\,mm$\times$0.3\,mm$\times$0.12\,mm.
	}	
	\label{fig:structure}
\end{figure}

For the study of orbital and electronic excitations in Mott insulators, RIXS and 
optical spectroscopy are complementary techniques. 
In the case of inversion symmetry, RIXS at the transition-metal $L$ edge is in particular 
sensitive to local, parity-conserving excitations between 
$d$ orbitals \cite{Ament11,Gretarsson19,Takahashi21,Yuan17,Kusch18,Nag18,Davies19}. 
In contrast, optical spectroscopy is most sensitive to electric-dipole-active 
absorption features such as inter-site excitations across the Mott gap or 
excitons \cite{Vergara22,Goessling08,Reul12}, 
while local excitations between $d$ orbitals mainly contribute 
in a phonon-assisted process, i.e., they become electric-dipole-active by the simultaneous
excitation of a symmetry-breaking phonon \cite{Henderson,Hitchman,Rueckamp05,Benckiser08}.
Interesting examples studied by both techniques are, e.g., the $3d^2$ vanadates $R$VO$_3$ 
\cite{Benckiser08,Benckiser13,Reul12} or the excitations from $J$\,=\,1/2 to 3/2 in the 
$4d^5$ Kitaev material $\alpha$-RuCl$_3$ \cite{Lebert19,Warzanowski20,Suzuki21}. 
In the infrared data, weak phonon-assisted absorption features can only be detected 
in the transparent range with excitation energies smaller than the Mott gap. 
In comparison, RIXS offers the possibility 
to observe the corresponding orbital excitations also for larger excitation energies 
but with limited energy resolution. 
At the Os $L_3$ edge with an incident energy $E_{\rm in}$\,=\,10.870\,keV, we achieve 
a resolution $\delta E$\,=\,63\,meV, i.e., $\delta E/E_{\rm in} \approx 6 \times 10^{-6}$. 
In contrast, the Fourier spectrometer employed for infrared measurements offers a resolution 
better than 0.1\,meV.\@ Accordingly, it is more difficult to determine the electronic parameters 
by using only one of the two techniques, as discussed, e.g., for RIXS on 
Yb$_2$Os$_2$O$_7$ \cite{Davies19} or infrared spectroscopy onK$_{\rm 2}$OsCl$_{\rm 6}$ \cite{Kozikowski83}. 
For K$_{\rm 2}$OsCl$_{\rm 6}$, we will show below that the combination of the 
two techniques resolves all ambiguities.

\section{Single-crystal growth and crystal structure}

Single crystals of K$_2$OsCl$_6$ and of the bromides $A_2$OsBr$_6$ with $A$\,=\,K and Rb 
were grown starting from commercially available dihydrogen hexahalogenoosmate(IV) H$_2$OsCl$_6$, 
resp.\ H$_2$OsBr$_6$ and the respective halide KCl or $A$Br. The educts 
were dissolved in stoichiometric ratio in diluted hydrochloric (resp.\ hydrobromic) acid. 
For the potassium compounds, single crystals of several mm size were achieved by slow, 
controlled evaporation of the solvent at 293\,K during a typical growth period of 
one to two weeks. For the Rb compound, crystals were of sub-mm size due to the low 
solubility. Examples of K$_{\rm 2}$OsBr$_{\rm 6}$ and K$_{\rm 2}$OsCl$_{\rm 6}$ 
are shown in Fig.\ \ref{fig:structure}.

At room temperature, these K$_2$PtCl$_6$-type compounds show a cubic crystal structure with 
space group $Fm\bar{3}m$ \cite{Fergusson74,Armstrong78,Saura22}, see Fig.\ \ref{fig:structure}. 
The Os$^{4+}$ ions are located on an \textit{fcc} lattice. 
The stoichiometry of all three compounds has been verified on small crystals by 
x-ray diffraction in the cubic phase \cite{Bertin23}.  
At 45\,K, K$_{\rm 2}$OsCl$_{\rm 6}$ exhibits a phase transition to a tetragonal phase \cite{Armstrong78}.
For K$_{\rm 2}$OsBr$_{\rm 6}$, the structural transition at 220\,K to a phase with tetragonal 
symmetry is followed by a phase transition to a monoclinic phase at 200\,K \cite{Saura22}.

\begin{figure}[t]
	\centering
	\includegraphics[width=\columnwidth]{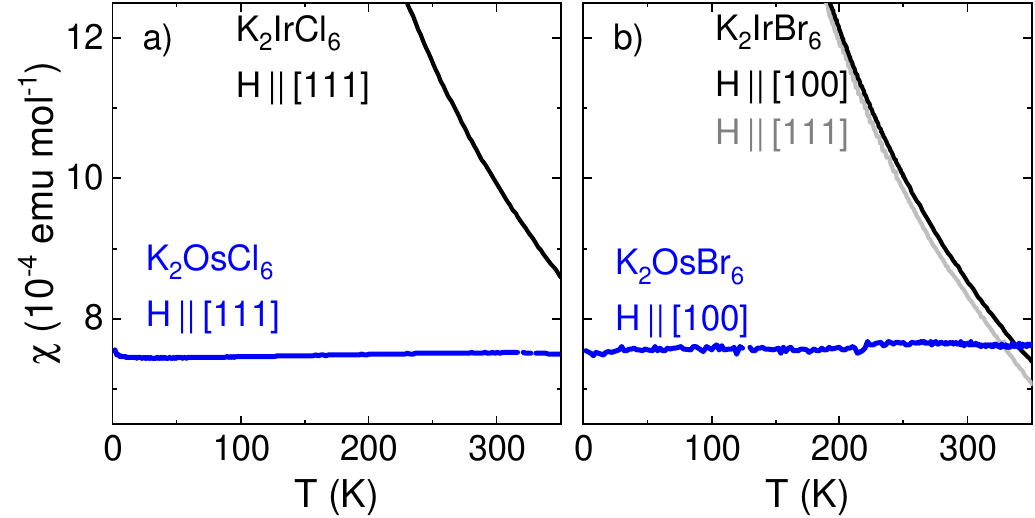}
	\caption{{\bf Magnetic susceptibility $\chi$ of K$_2$OsCl$_6$ and K$_2$OsBr$_6$.}
		Both compounds show dominant temperature-independent Van-Vleck paramagnetism. 
		For comparison, we show $\chi$ of the $J$\,=\,1/2 iridium sister compounds with a 
		sizable Curie-Weiss contribution.
		\label{fig:Susceptibility}
	}
\end{figure}

\section{Magnetic susceptibility}

The magnetic susceptibility $\chi$ of K$_{\rm 2}$OsCl$_{\rm 6}$ and K$_{\rm 2}$OsBr$_{\rm 6}$ 
is plotted in Fig.\ \ref{fig:Susceptibility}. 
We observe a temperature-independent susceptibility without any indication of a Curie-Weiss contribution 
of defects. The constant value $\chi$\,=\,$7.5\cdot 10^{-4}$\,emu/mol can be attributed to a dominant 
Van-Vleck term and a small diamagnetic contribution of the closed shells of K$^+$, Os$^{4+}$, Cl$^-$,  
and Br$^-$, $\chi$\,=\,$\chi_{VV} + \chi_{\rm dia}$.
Via tabulated values \cite{Bain08}, we estimate
$\chi_{\rm dia}^{\rm Cl}$\,=\,$-2.1 \cdot 10^{-4}$\,emu/mol and 
$\chi_{\rm dia}^{\rm Br}$\,=\,$-2.7 \cdot 10^{-4}$\,emu/mol. 
This yields experimental values of the Van-Vleck contribution of 
$\chi_{VV}^{\rm Cl}$\,=\,$9.6 \cdot 10^{-4}$\,emu/mol and 
$\chi_{VV}^{\rm Br}$\,=\,$10.2 \cdot 10^{-4}$\,emu/mol.

Our spectroscopic data establish a $J$\,=\,0 ground state of the 
$5d^4$ configuration of Os$^{4+}$, cf.\ Sect.\ \ref{sect:resultsCl}.  		
The corresponding Van-Vleck susceptibility is given by
\begin{equation}
	\chi_{\rm VV} = \frac{N}{V} 2 \mu_B^2  \sum_n \frac{\left| \langle n|L_z + g S_z |0\rangle \right|^2}{E_n - E_0} \, , 
\end{equation}
where $|0\rangle$ and $|n\rangle$ denote the ground state and excited states, respectively, 
$E_n-E_0$ is the respective energy difference, $S_z$ and $L_z$ refer to the $z$ components of 
spin and orbital angular momentum, $\mu_B$ is the Bohr magneton, and $N/V$ denotes the density 
of Os ions. In the most simple picture, one can consider only the matrix elements from the 
$J$\,=\,0 ground state to the lowest excited state, the $J$\,=\,1 triplet, assuming that 
all other excitation energies are infinite. 
The corresponding matrix element equals $\left| \langle 1|L_z + g S_z |0\rangle \right|^2$\,=\,6. 
With an excitation energy of 0.35\,eV (see below), this yields 
$\chi_{VV}^{J=1}$\,=\,$11\cdot 10^{-4}$ emu/mol. 
As a more sophisticated alternative, we calculate the matrix elements via 
\textit{Quanty} \cite{Haverkort12,Haverkort16}, 
using the electronic parameters that result from our thorough analysis of the spectroscopic 
data discussed below. For the 209 possible excited states of a $5d^4$ configuration, 
we find that only two states contribute significantly. About 90\,\% arise from the $J$\,=\,1 
state at about 0.35\,eV, while 10\,\% stem from a state with one electron in the $e_g$ 
orbitals at about 3.5\,eV.\@ The two matrix elements are similar, hence the contribution 
of the second term is suppressed by about a factor 10 due to the higher excitation energy. 
We find $\chi_{VV,mod}^{\rm Cl}$\,=\,$8.8\cdot 10^{-4}$ emu/mol 
and $\chi_{VV,mod}^{\rm Br}$\,= $8.6\cdot 10^{-4}$\,emu/mol, 
in reasonable agreement with the experimental result.

\section{Spectroscopic measurements}
\label{sec:spectroscopy}

All RIXS experiments were performed at beamline ID20 of the European Synchrotron Radiation 
Facility. Incident photons from three consecutive U26 undulators were monochromatized by a 
Si(111) high-heat-load monochromator and either a successive Si(664) backscattering-channel-cut 
or a Si(311) channel-cut post-monochromator at 10.870\,keV with a final bandwidth of 18\,meV 
or 0.29\,eV, respectively. The monochromatic X-ray beam was focused by a mirror system in 
Kirkpatrick-Baez geometry to $8 \times 50$\,$\mu$m$^2$ (V $\times$ H) at the sample position. 
Incident $\pi$ polarization in the horizontal scattering plane was used. 
We specify the transferred momentum $\mathbf{q}$ in reciprocal lattice units. 
First, we studied the resonance behavior of K$_2$OsCl$_6$ at the Os $L_3$ edge by measuring 
low-energy-resolution RIXS spectra with the incident energy in the range from 10.866 to 
10.880\,keV.\@ Then, the incident energy was tuned to 10.870\,keV to maximize the 
intra-$t_{2g}$ excitations. 
RIXS spectra were measured using the 2\,m analyzer/detector arm of the spectrometer. 
The Si(6,6,4) reflection of a diced Si(5,5,3) analyzer crystal was utilized in conjunction 
with a pixelated area detector \cite{Huotari2005,Huotari2006,Moretti2013}. The overall 
energy resolution of the setup was 63\,meV for the high-energy-resolution spectra and 
0.29\,eV for the low-energy-resolution spectra, respectively, as estimated by the full width 
at half maximum of quasielastic scattering from a piece of adhesive tape. 
To determine the energy-loss scale of the spectrometer, we first define its origin at the 
center of mass of the rocking curve of a diced Si(664) analyzer crystal using quasielastic 
scattering from a piece of adhesive tape. Then, the increment of the scale is determined 
mainly by the analyzer Bragg angle and detector position. 
The combination of RIXS and optics allows us to examine the accuracy of 
this approach up to high energies. Comparing the excitation energy of a RIXS peak with 
the corresponding feature in the optical data at 2117\,meV (see below), we find that 
the two values agree within about 1\,\%. This excellent result is in line with a previous 
study of the precision of the RIXS energy scale for energies up to 150\,meV \cite{Moretti18}. 
For a consistent analysis, we have anchored the RIXS energy loss scale using the 
optical value of 2117\,meV.\@
The RIXS measurements were performed using a dynamic helium gas flow cryostat as 
described elsewhere \cite{vanderLinden2016}.
RIXS data were collected at 20\,K and 300\,K on a (111) surface, with (001) and (110) lying 
in the horizontal scattering plane. 
All RIXS spectra are corrected for energy-dependent self absorption \cite{Minola15}.

Infrared transmittance measurements in the energy range from 0.1 to 2.5\,eV were 
performed using a Bruker IFS 66/v Fourier-transform spectrometer equipped with a 
continuous-flow $^{4}$He cryostat. 
The transmittance $T(\omega)$ was measured at several temperatures in the range from 
5 to 300\,K.\@ We employed thin plane-parallel samples with a cubic (111) surface. 
The sample of K$_{\rm 2}$OsCl$_{\rm 6}$ was lapped to a suitable thickness and polished 
with CeO$_2$ in propanol, while the sample of K$_{\rm 2}$OsBr$_{\rm 6}$ was measured as grown.
The sample thickness amounts to $d$\,=\,120(5)\,$\mu$m for K$_2$OsCl$_6$ and 
$d$\,=\,170(7)\,$\mu$m for K$_2$OsBr$_6$.
We determined the complex optical conductivity 
$\sigma(\omega)$\,=\,$\sigma_1(\omega) + i \sigma_2(\omega)$ from the complex index of 
refraction $n(\omega) +i\kappa(\omega)$, which, in turn, can be derived from $T(\omega)$. 
The imaginary part $\kappa(\omega)$ depends sensitively on the  absolute value of $T(\omega)$. 
In the transparent range with $\kappa \ll n$, the real part $n(\omega)$ is nearly constant 
and can be obtained from the period of the Fabry-P\'erot interference fringes 
which arise from multiple reflections within the sample. 
To study the optical response in the non-transparent range above the Mott gap, 
we performed ellipsometry measurements on K$_2$OsCl$_6$ at 300\,K 
in the range from 1 to 6\,eV using a Woollam VASE ellipsometer.

\section{Results on K$_2$O\lowercase{s}C\lowercase{l}$_6$}
\label{sect:resultsCl}

\subsection{Resonance behavior}

\begin{figure}[t]
	\centering
	\includegraphics[width=\columnwidth]{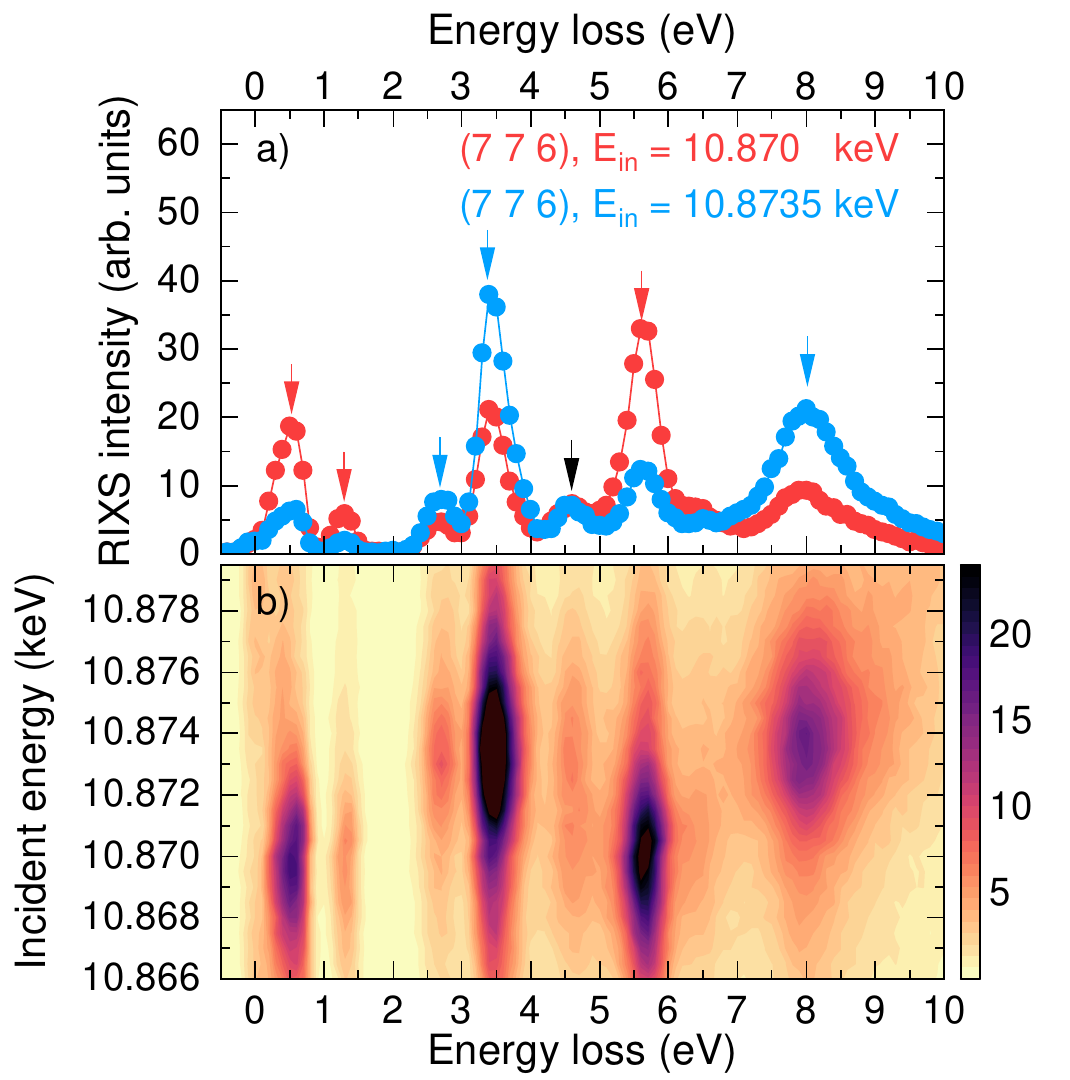}
	\includegraphics[width=\columnwidth]{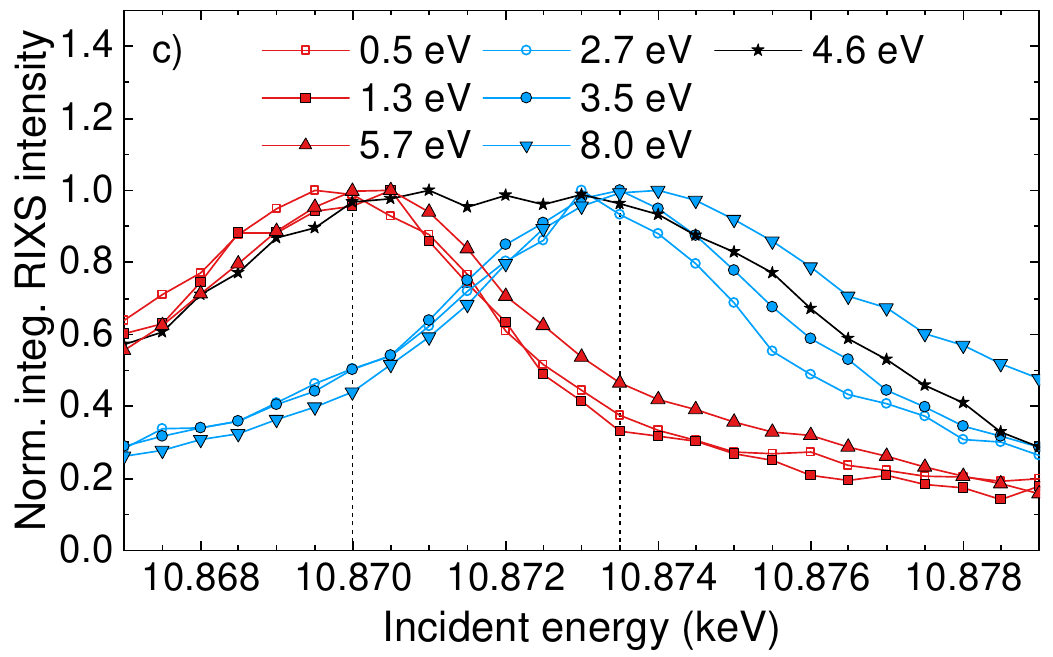}
	\caption{{\bf Resonance behavior of K$_{\rm 2}$OsCl$_{\rm 6}$ at $T$\,=\,20\,K.}  
	a) Low-resolution RIXS spectra for incident energy $E_{\rm in}$\,=\,10.870\,keV 
	and 10.8735\,keV, i.e., at $t_{2g}$ and $e_g$ resonance, respectively. 
	Arrows indicate peak energies and their color the resonance behavior. 
	b) Low-resolution resonance map of the RIXS intensity based on RIXS spectra 
	measured with different $E_{\rm in}$ for $\mathbf{q}$\,=\,(7\,\,7\,\,6).
    c) Normalized integrated intensity of the RIXS peaks as a function of $E_{\rm in}$. 
	The integration interval is chosen to be $\pm$0.4\,eV around the peak energy. 
	With the exception of the 4.6\,eV peak, all peaks show either $t_{2g}$ or 
	$e_g$ resonance.
}
\label{fig:map}
\end{figure}

To study the resonance behavior and to maximize the RIXS intensity, we collected 
low-resolution RIXS spectra of K$_2$OsCl$_6$ for different incident energies at $T$\,=\,20\,K 
for a transferred momentum $\mathbf{q}$\,=\,(7\,\,7\,\,6), see Fig.\,\ref{fig:map}b).
With the energy loss being independent of $E_{\rm in}$ for all of the observed RIXS peaks, 
we find the two distinct resonance energies $E_{\rm in}$\,=\,10.870\,keV and 10.8735\,keV, 
as shown in Fig.\,\ref{fig:map}c).
These can be attributed to $t_{2g}$ resonance and $e_g$ resonance, i.e., 
resonance enhancement of the initial x-ray absorption part of the RIXS process in which 
a $2p$ core electron is promoted to either a $t_{2g}$ or an $e_g$ orbital, respectively. 
The difference of the two resonance energies gives a first estimate of the cubic 
crystal-field splitting 10\,$Dq$\,$\approx$\,3.5\,eV.\@ This agrees with the strong RIXS 
peak at about 3.5\,eV energy loss that exhibits $e_g$ resonance behavior and corresponds 
to an excitation from a $t_{2g}$ orbital to an $e_g$ orbital. 
Cuts of the resonance map at the two resonance energies are shown in Fig.\,\ref{fig:map}a).

\subsection{Character of RIXS features}

Based on the resonance behavior, we distinguish three different kinds of excitations: 
intra-$t_{2g}$ excitations, crystal-field excitations to $e_g$ orbitals, and charge-transfer excitations. 
At low energy loss, up to about 2\,eV, all RIXS features exhibit $t_{2g}$ resonance 
and can be attributed to intra-$t_{2g}$ excitations. Their excitation energies mainly 
reflect spin-orbit coupling $\zeta$ and Hund's coupling $J_H$. 
These intra-$t_{2g}$ excitations will be addressed below by high-resolution RIXS 
measurements and optical transmission measurements. 
At higher energy loss, the RIXS peaks in the range from 2.7 to 4.6\,eV are resonantly 
enhanced for $E_{\rm in}$\,=\,10.8375\,keV and correspond to crystal-field excitations to 
$e_g$ orbitals, $|t_{2g}^4\rangle \rightarrow |t_{2g}^3\,e_g^1\rangle$. 
We will show below that the lowest $e_g$ excitation at 2.7\,eV can be identified with 
the high-spin $S$\,=\,2 multiplet with $^5E$ symmetry. The energy of this high-spin state 
is reduced by Hund's coupling $J_H$, and the $^5E$ multiplet becomes the $d^4$ ground state 
if $J_H$ dominates over the cubic crystal-field splitting 10\,Dq.

At still higher energy loss, we observe charge-transfer excitations, 
$|5d_{\rm Os}^4 \, 3p_{\rm Cl}^6 \rangle$\,$\rightarrow$\,$|5d_{\rm Os}^5 3p_{\rm Cl}^5 \rangle$. 
The peaks at 4.6\,eV and 5.7\,eV both show $t_{2g}$ resonance and hence can 
be attributed to $|t_{2g}^5 \, 3p_{\rm Cl}^5 \rangle$ final states. 
Roughly, these two excited states with $t_{2g}^5$ configuration can be identified with 
$J$\,=\,1/2 and 3/2 on the Os site. 
The peak at 8.0\,eV energy loss shows $e_g$ resonance and corresponds to excitations 
from Cl $3p$ to Os $5d$ $e_g$ orbitals. 
The RIXS peak at 4.6\,eV is the only one that resonates both at 10.870\,keV and at 
10.8735\,keV, see lower panel of Fig.\,\ref{fig:map}. This suggests an overlap between 
an on-site crystal-field excitation to $e_g$ orbitals and the lowest charge-transfer 
excitation to $t_{2g}$ orbitals.
Note that the energy difference of about 3.4\,eV between the peaks at 8.0\,eV and 4.6\,eV 
confirms our first rough estimate of 10\,$Dq$\,$\approx$\,3.5\,eV.\@ 
Accordingly, we identify the peak energy of the lowest charge-transfer excitation with 
the charge-transfer energy, $\Delta_{\rm CT}$\,=\,4.6\,eV.

The occurrence of both $t_{2g}$ resonance and $e_g$ resonance behavior of 
charge-transfer excitations has previously been observed in, e.g., 
$5d^5$ K$_2$IrBr$_6$ \cite{ReigiPlessis20}, $5d^4$ Yb$_2$Os$_2$O$_7$ \cite{Davies19}, 
and $5d^2$ Ba$_2$YReO$_6$ \cite{Yuan17}. 
In the present case of K$_{\rm 2}$OsCl$_{\rm 6}$, the charge-transfer excitations are particularly 
well-defined and yield comparably narrow RIXS peaks. For instance 
the 5.7\,eV peak shows a full width at half maximum of 0.6\,eV.\@ 
This suggests that the charge-transfer excitations to $t_{2g}$ orbitals are 
well localized in non-magnetic K$_{\rm 2}$OsCl$_{\rm 6}$, in agreement with the notion that interactions 
between the well separated OsCl$_6$ octahedra are very small, at least for $t_{2g}$ orbitals.

\subsection{Intra-$t_{2g}$ excitations of K$_2$OsCl$_6$ in RIXS}
\label{sec:intrat2g}

\begin{figure}[t]
	\centering
	\includegraphics[width=\columnwidth]{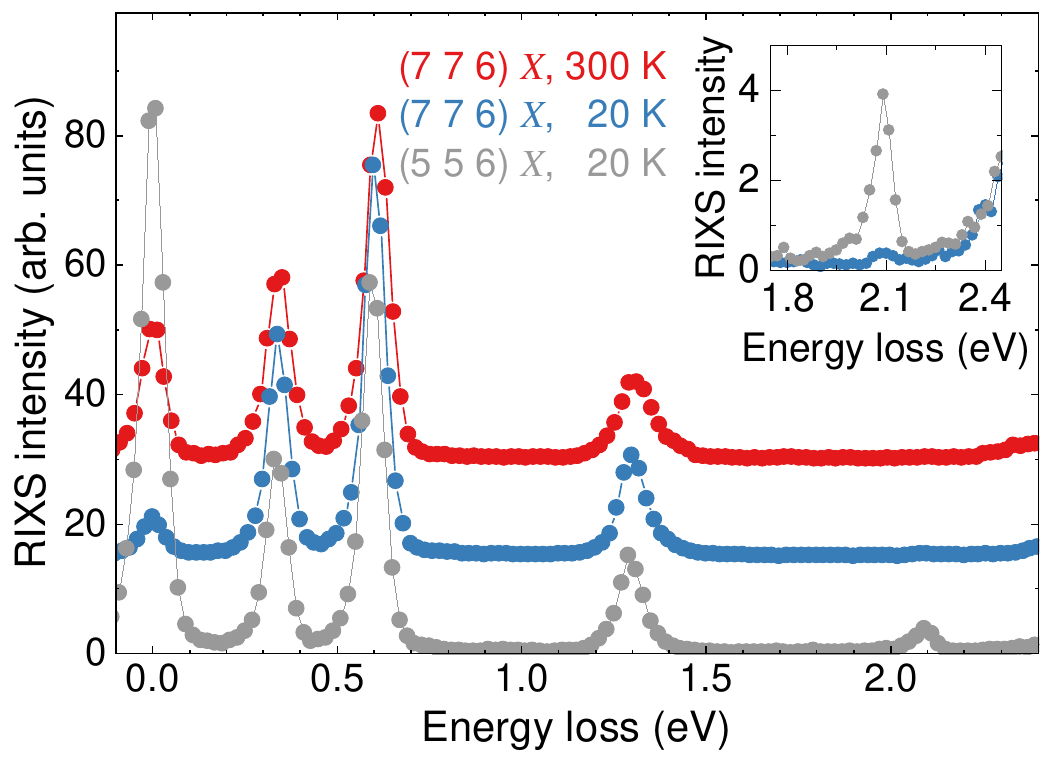}
	\caption{{\bf Intra-$t_{2g}$ excitations of K$_{\rm 2}$OsCl$_{\rm 6}$ at the $X$ point.}
		RIXS spectra at $\mathbf{q}$\,=\,(5\,\,5\,\,6) and (7\,\,7\,\,6) reveal four peaks at 
		0.3, 0.6, 1.3, and 2.1\,eV with little temperature dependence. 
		For clarity, the data are plotted with an offset. The inset highlights 
		the $^1\!A_1$ peak at 2.1\,eV which is suppressed for a scattering 
		angle $2\theta$\,=\,90$^\circ$, see main text. 
		\medskip
		\label{fig:t2g_RIXS_IR}
	}
	%
	\centering
	\includegraphics[width=\columnwidth]{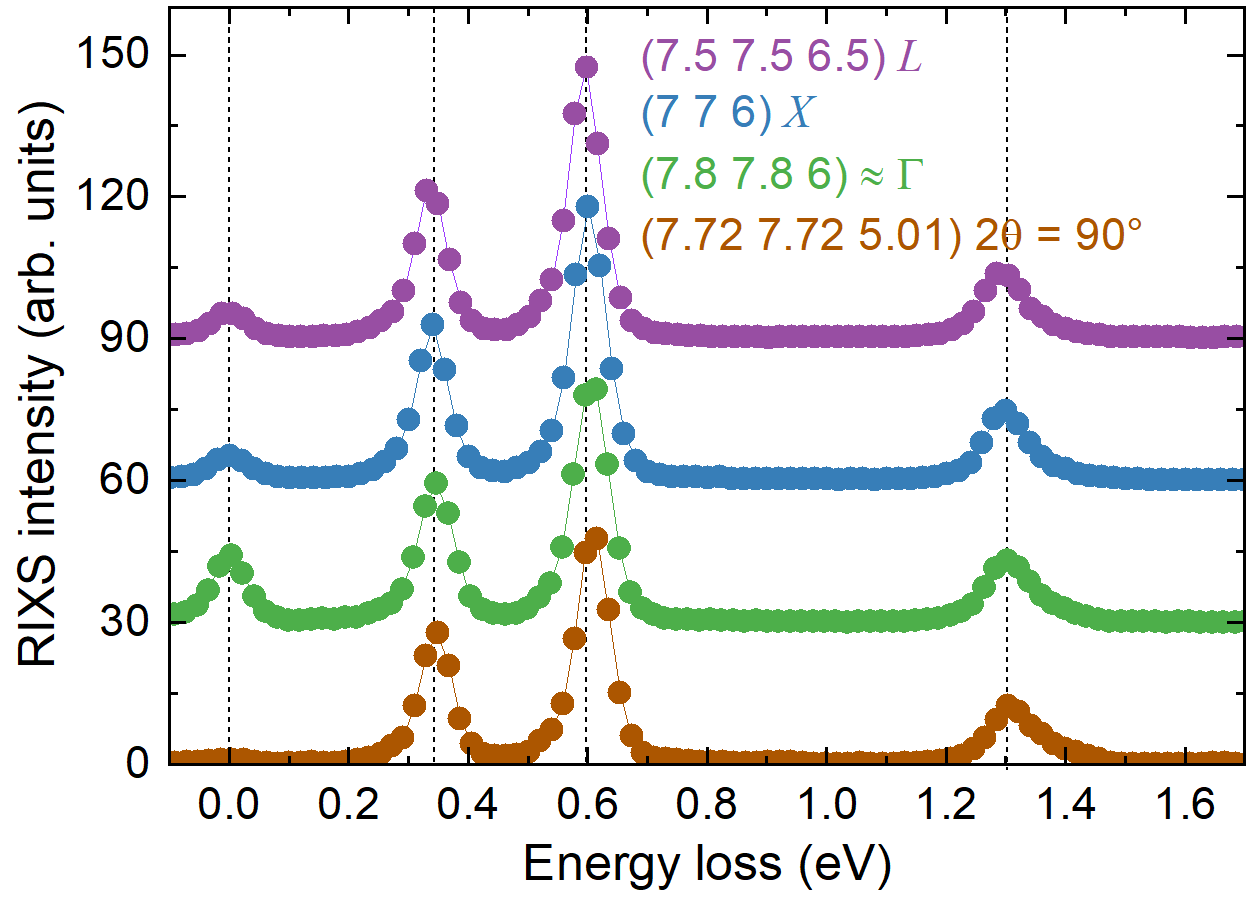}
	\caption{{\bf Low-energy RIXS features in K$_{\rm 2}$OsCl$_{\rm 6}$ at 20\,K at $L$, $X$, and close to $\Gamma$.} 
		The latter data were measured in the vicinity of $\Gamma$ to avoid the strong contribution 
		of Bragg scattering right at $\Gamma$. 	Dashed lines indicate peak positions at $X$. 
		The dispersion is negligible, highlighting the local character of the intra-$t_{2g}$ excitations. Additionally, data at (7.72\,\,7.72\,\,5.01) demonstrate the suppression 
		of elastic Thomson scattering for 2$\theta$\,=\,90$^\circ$. 
		\label{fig:dispersion}
	}
\end{figure}

High-resolution RIXS spectra of the intra-$t_{2g}$ excitations reveal four peaks at 
about 0.3, 0.6, 1.3, and 2.1\,eV, see Fig.\,\ref{fig:t2g_RIXS_IR}. 
The local, on-site character of these excitations is highlighted by the insensitivity 
of the peak energies to the transferred momentum $\mathbf{q}$, i.e., the absence of a 
measurable dispersion, see Fig.\ \ref{fig:dispersion}. 
For a first assignment, we neglect spin-orbit coupling, $\zeta$\,=\,0, and assume 
10\,$Dq$\,=\,$\infty$ for the cubic crystal-field splitting. In this case,
the effect of the $e_g$ orbitals on the intra-$t_{2g}$ excitations vanishes. 
In a $d^4$ configuration there are 15 $t^4_{2g}$ states. Inter-orbital Coulomb 
interactions lift their degeneracy and yield a nine-fold degenerate $^3T_1$ ground state 
and the $^1T_2$, $^1E$, and $^1\!A_1$ excited states. The latter is expected 
at $5J_H$, while $^1T_2$ and $^1E$ show an energy of $2J_H$ \cite{Georges13,Zhang17}. 
Note, however, that these values correspond to a $t_{2g}$-only scenario, 
i.e.\ 10\,$Dq$\,=\,$\infty$, 
they are not suitable to estimate $J_H$ for realistic values of 10\,$Dq$ and $\zeta$, 
as discussed below.
The low-energy RIXS peaks at 0.3\,eV and 0.6\,eV reflect strong spin-orbit coupling, 
which splits the $^3T_1$ multiplet into the $J$\,=\,0 ground state 
with $\Gamma_1$ symmetry and two excited states 
of $J$\,=\,1 ($\Gamma_4$) and $J$\,=\,2 ($\Gamma_3 + \Gamma_5$) character, respectively. 
The two contributions to the $J$\,=\,2 peak at about 0.6\,eV are expected 
to split in energy for finite values of the cubic crystal-field splitting 10\,$Dq$. 
The fact that our RIXS data still show a single peak, cf.\ Fig.\ \ref{fig:dispersion}, 
indicates that the splitting is much smaller than the energy resolution $\delta E$\,=\,63\,meV.\@ 
A finite Coulomb-induced splitting of the $J$\,=\,2 states is most relevant in 
cubic $5d^2$ compounds. In these electron analogs of the two-$t_{2g}$-hole $5d^4$ 
configuration, the $J$\,=\,2 multiplet has the lowest energy and its splitting for finite 
$10\,Dq$ determines the local ground state, which may lead to exotic 
multipolar phases \cite{Paramekanti20,Lovesey20,Khaliullin21,Voleti21,Pourovskii21,Rayyan23}.
For both $5d^4$ and $5d^2$, also the degeneracy of the $^1T_2 $ and $^1E$ multiplets 
is lifted for finite 10\,$Dq$. 
The asymmetric line shape of the RIXS peak around 1.3\,eV at 20\,K and 300\,K, 
cf.\ Fig.\ \ref{fig:t2g_RIXS_IR}, indeed indicates a finite splitting between the 
cubic multiplets with a weaker feature on the high-energy side, 
as discussed in K$_2$RuCl$_6$ and Ba$_2$YIrO$_6$ \cite{Takahashi21,Kusch18}.

\begin{figure}[t]
	\centering
	\includegraphics[width=\columnwidth]{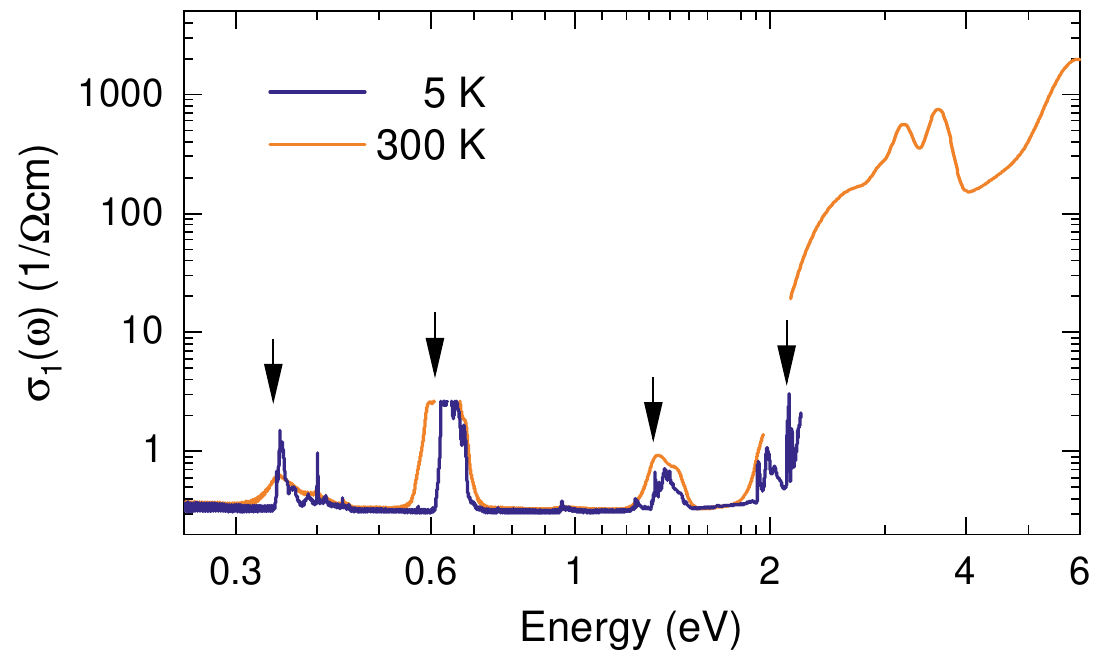}
	\caption{{\bf Optical conductivity $\sigma_1(\omega)$ of K$_{\rm 2}$OsCl$_{\rm 6}$.}
		Note the logarithmic scales. The strong absorption features above the Mott gap at 
		2.2\,eV have been measured by ellipsometry, while the data in the transparent range 
		below the gap with much smaller values of $\sigma_1(\omega)$ are based on the 
		transmittance $T(\omega)$. 
		The latter is suppressed below the noise level above 2.2\,eV and around 0.6\,eV, 
		limiting the maximum value of $\sigma_1(\omega)$ that can be determined via $T(\omega)$ 
		for the given sample thickness.
		Below the Mott gap, $\sigma_1(\omega)$ shows (phonon-assisted) intra-$t_{2g}$ 
		excitations in agreement with RIXS (arrows depict the RIXS intra-$t_{2g}$ peak energies).
		Additionally, $\sigma_1(\omega)$ reveals an exciton around 2.0\,eV and two tiny features 
		at about 0.95 and 1.25\,eV that can be assigned to inter-site overtones of the 
		low-energy intra-$t_{2g}$ excitations. 
		\label{fig:IR_elli}
	}
\end{figure}

Overall, the excitation spectrum agrees with previous RIXS results on $5d^4$ iridates 
\cite{Yuan17,Kusch18,Nag18,Aczel22}. 
RIXS data of Yb$_2$Os$_2$O$_7$ \cite{Davies19} also show 
the two low-energy modes below 1\,eV but the intra-$t_{2g}$ features at higher energies 
are hidden by a broad band that has been attributed to defects. 
Furthermore, the energy of the lowest excitation is much smaller in 
Yb$_2$Os$_2$O$_7$ due to a trigonal distortion \cite{Davies19}.
In $4d^4$ K$_2$RuCl$_6$, the equivalent of the 
lower three features have been observed \cite{Takahashi21}. The $^1\!A_1$ peak at 2.1\,eV 
thus far was only observed as a very weak feature in Sr$_2$YIrO$_6$ \cite{Yuan17}. 
Comparing two data sets measured at the equivalent $X$ points (5\,\,5\,\,6) and (7\,\,7\,\,6), 
see Fig.\ \ref{fig:t2g_RIXS_IR}, 
we find that the $^1\!A_1$ peak is suppressed in the latter, which has been recorded with 
a scattering angle $2\theta$ close to 90$^\circ$. This geometry typically is chosen to 
suppress elastic Thomson scattering, explaining the absence of this peak in previous 
measurements on $5d^4$ compounds. 
An example for the successful suppression of the elastic line for 
$2\theta$\,=\,90$^\circ$ is given by the data for (7.72\,\,7.72\,\,5.01) in 
Fig.\ \ref{fig:dispersion}. In contrast, the data for (5\,\,5\,\,6) were measured with 
$2\theta$\,=\,66$^\circ$ and accordingly show a more pronounced elastic line. 
For $2\theta$\,=\,90$^\circ$, the polarization of the scattered light is perpendicular to 
the incident $\pi$ polarization. This suggests that the $^1\!A_1$ feature 
is observable only for parallel polarization, which is supported by simulations using 
\textit{Quanty} \cite{Haverkort12,Haverkort16}.

Our peak assignment assumes cubic symmetry and neglects the phase transition 
from cubic to tetragonal symmetry at 45\,K.\@ 
However, the RIXS data measured at 20\,K and 
300\,K fully agree with each other, see Fig.\ \ref{fig:t2g_RIXS_IR}, suggesting that the 
non-cubic splitting is much smaller than the energy resolution $\delta E$\,=\,63\,meV.\@ 
This is supported by the infrared data, which were measured with $\delta E$\,=\,0.25\,meV.\@ 
A thorough analysis of the temperature dependence of the infrared data reveals a non-cubic 
splitting of about 4\,meV, as discussed below.

\subsection{Optical conductivity of K$_2$OsCl$_6$}

\begin{figure}[t]
	\centering
	\includegraphics[width=\columnwidth]{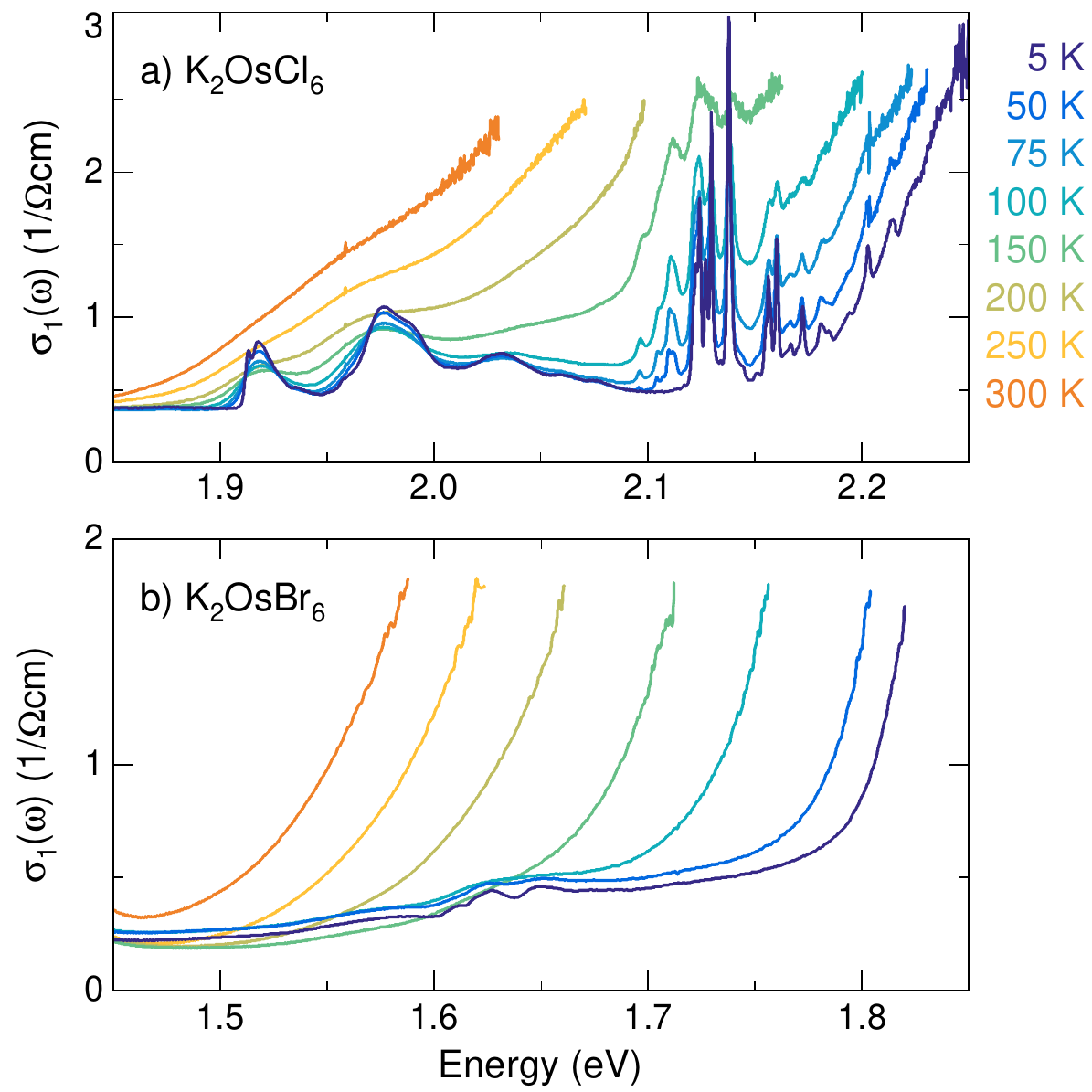}
	\caption{{\bf Temperature dependence of $\sigma_1(\omega)$ close to the Mott gap 
	in a) K$_2$OsCl$_6$ and b) K$_2$OsBr$_6$.}
	In K$_2$OsCl$_6$, the onset of excitations across the Mott gap is observed 
	at 2.2\,eV at 5\,K.\@	
	This onset is washed out with increasing temperature since additional processes with 
	simultaneous annihilation of thermally excited phonons reduce the total excitation 
	energy in the Urbach tail. The $^1\!A_1$ intra-$t_{2g}$ excitation is observed at 
	2.117\,eV, and the many narrow lines around it correspond to phonon sidebands. 
	In contrast to the $^1\!A_1$ peak, the absorption band between 1.9 and 2.1\,eV has 
	no counterpart in RIXS and can be assigned to an exciton. 
	For comparison, the data of K$_2$OsBr$_6$ 
	show the Mott gap at 1.8\,eV at 5\,K and a very similar temperature dependence. 
	In K$_2$OsBr$_6$, the exciton is observed around 1.6-1.7\,eV.\@
}
\label{fig:exciton}
\end{figure}

Based on the different selection rules, optical spectroscopy and RIXS are complementary 
techniques. The optical conductivity $\sigma_1(\omega)$ is dominated by 
electric-dipole-active transitions. Considering the orbital and electronic excitations 
in a Mott insulator, the dominant absorption features arise from \textit{inter-site} processes 
such as excitations across the Mott gap, 
in our case $|d^4_i\,d^4_j\rangle$\,$\rightarrow$\,$|d^3_i\,d^5_j\rangle$, 
or charge-transfer excitations 
$|5d_{\rm Os}^4\,3p_{\rm Cl}^6\rangle$\,$\rightarrow$\,$|5d_{\rm Os}^5\,3p_{\rm Cl}^5\rangle$. 
In Fig.\ \ref{fig:IR_elli}, the Mott gap corresponds to the steep rise of $\sigma_1(\omega)$ 
above 2.2\,eV.\@ 
In contrast, RIXS at the Os $L$ edge is dominated by on-site excitations \cite{Ament11,Yuan17,Kusch18,Nag18,Davies19}, 
as discussed above, in particular for incident energies tuned to $t_{2g}$ resonance.  
In the presence of inversion symmetry, on-site crystal-field excitations such as the 
intra-$t_{2g}$ excitations are parity forbidden in $\sigma_1(\omega)$ but can become 
weakly allowed for instance in a phonon-assisted process. 
The corresponding spectral weight is several orders of magnitude smaller. 
Such weak features can only be studied for energies below the Mott gap where they are 
not hidden by stronger absorption processes, see Fig.\ \ref{fig:IR_elli}.

\subsubsection{Intersite excitations in $\sigma_1(\omega)$}

Concerning the strong absorption features in $\sigma_1(\omega)$ above the Mott gap, 
we have to distinguish charge-transfer excitations between Cl and Os states 
from Mott-Hubbard excitations between Os states on different sites. 
One expects a larger spectral weight for the charge-transfer excitations due to the 
larger hopping matrix elements between Cl $3p$ and Os $5d$ states compared to Os 
inter-site hopping. Accordingly, the strongest peak at about 6\,eV can be assigned to 
a charge-transfer excitation, in agreement with the RIXS feature at 5.7 eV.\@ 
The three peaks in $\sigma_1(\omega)$ between 2 and 4\,eV correspond to 
Mott-Hubbard excitations involving $t_{2g}$ states, 
$|{t_{2g}^4}_i \, {t_{2g}^4}_j\rangle$\,$\rightarrow$\,$|{t_{2g}^3}_i\,{t_{2g}^5}_j\rangle$. 
Their energy mainly reflects the on-site Coulomb repulsion $U$, while the splitting 
is caused by $J_H$ and $\zeta$. In $4d^4$ Ca$_2$RuO$_4$, the relative spectral weight 
of these bands has been employed to estimate $\zeta$ \cite{Vergara22}. 
Comparing $\sigma_1(\omega)$ with the RIXS data, we emphasize that the origin of the 
RIXS features between 2 and 4\,eV is very different. RIXS shows on-site excitations 
to $e_g$ states, as demonstrated by the resonance behavior. 
The pronounced difference between the two techniques is based 
on the on-site energy $U$ that has to be paid in the optical Mott-Hubbard excitations.  
In $\sigma_1(\omega)$, Mott-Hubbard excitations to $e_g$ states 
$|{t_{2g}^4}_i \, {t_{2g}^4}_j\rangle$\,$\rightarrow$\,$|{t_{2g}^3}_i\,{(t_{2g}^4 e_g^1)}_j\rangle$
are expected to occur roughly 10\,$Dq$ higher in energy than the corresponding $t_{2g}$ bands, 
i.e., above about 6\,eV.\@ In K$_2$OsCl$_6$, these processes hence overlap with 
the charge-transfer excitations discussed above.

The onset of excitations across the Mott gap can be determined very well from 
the transmittance $T(\omega)$ which is strongly suppressed by the steep increase 
of absorption. This limits the accessible frequency window in our measurement on a 
single crystal with thickness  $d$\,=\,120\,$\mu$m, see Fig.\ \ref{fig:IR_elli}. 
At 5\,K, the onset is at $\Delta$\,=\,2.2\,eV.\@ With increasing temperature, 
the onset shifts to lower energy, roughly to 1.9\,eV at 300\,K, while the slope 
of the steep increase in $\sigma_1(\omega)$ is reduced, see Fig.\ \ref{fig:exciton}a).
The change of slope shows that the main origin of this shift is not a possible small 
temperature dependence of the gap itself. 
The enhanced absorption below 2.2\,eV predominantly can be attributed to thermally 
activated phonon-assisted excitations across the gap, i.e., the Urbach tail.

\begin{figure}[t]
	\centering
	\includegraphics[width=\columnwidth]{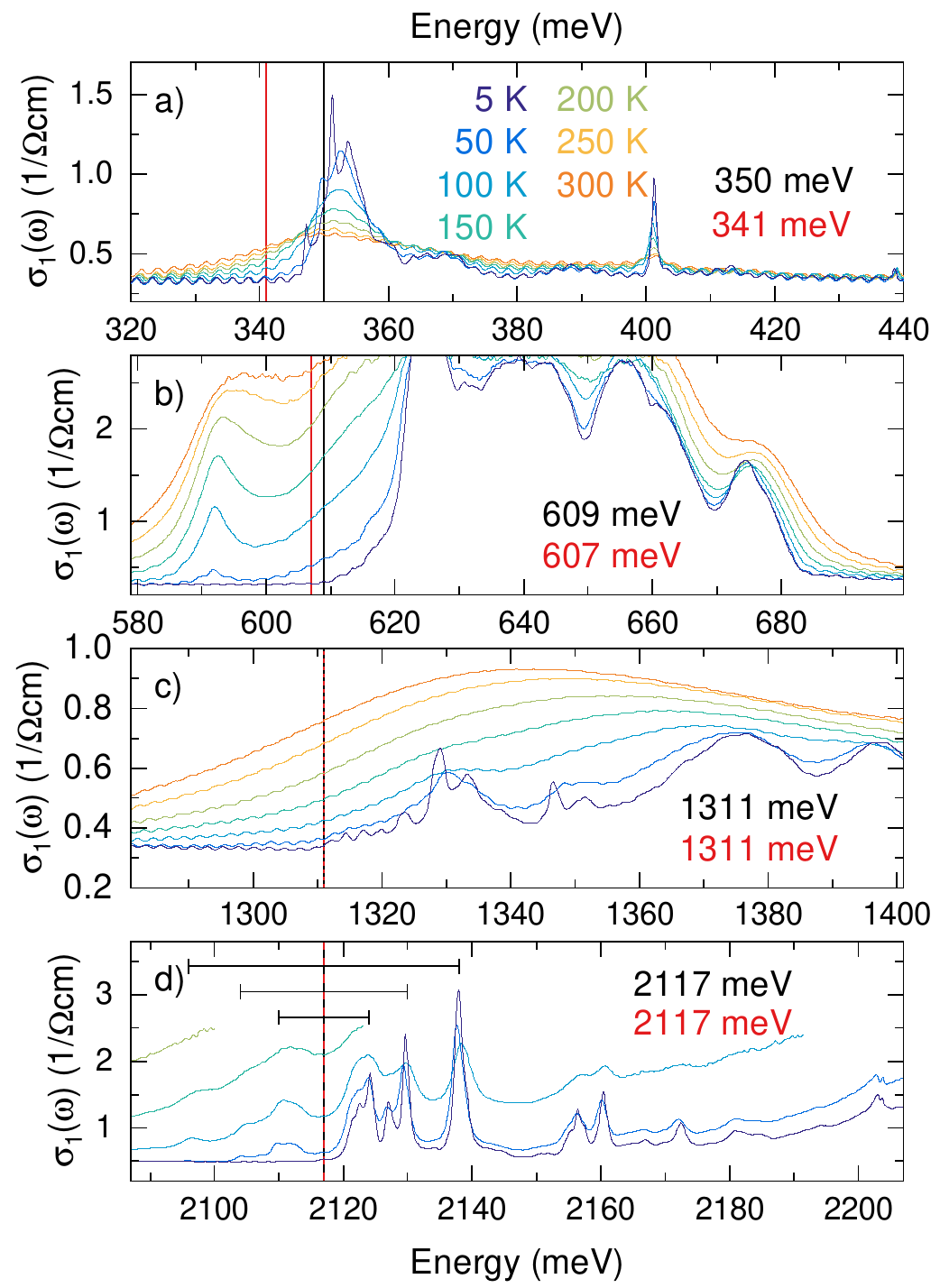}
	\caption{{\bf Intra-$t_{2g}$ excitations of K$_2$OsCl$_6$ in $\sigma_1(\omega)$.} 
     The panels depict excitations from $J$\,=\,0 to 
     a) $J$\,=\,1, i.e, $\Gamma_1$ to $\Gamma_4$; 
     b) the split $J$\,=\,2 states, i.e., $\Gamma_1$ to $\Gamma_5$ and $\Gamma_3$; 
     c) the split $^1T_2$ and $^1E$ states; and d) $^1\!A_1$. 
     Each panel covers a window of the same width, 120\,meV.\@ 		
  	 In each panel, the vertical black line denotes the bare electronic (zero-phonon) energy, 
     while the red line shows the fit result of the RIXS data. 
     In $\sigma_1(\omega)$, the spectral weight is dominated by phonon sidebands.
     In b), the transmittance around the peak maximum is suppressed below the noise level, 
     which limits the measurable range of $\sigma_1(\omega)$. 
     In d), the horizontal black lines denote phonon sidebands at 
     $E_{\,^1\!A_1} \pm  E_{\rm ph}$ with $E_{ph}$\,=\,7, 13, and 21\,meV.
}
	\label{fig:sigma_bands}
\end{figure}

Below the Mott gap, the optical data show three additional absorption bands that are 
absent in the RIXS data. The strongest one is observed between 1.9 and 2.1\,eV, 
which is very close to the Mott gap, see Fig.\ \ref{fig:exciton}a).
This feature is well separated from the $^1A_1$ intra-$t_{2g}$ excitation at 2.117\,eV.\@ 
Its spectral weight is comparable to the weak intra-$t_{2g}$ excitations but the absence 
of a corresponding RIXS feature strongly points to a different origin. 
We therefore assign it to an exciton with $5d^3$ and $5d^5$ configurations on 
neighboring Os sites. This exciton is stabilized by nearest-neighbor Coulomb attraction 
and may induce a local relaxation of the Cl$_6$ octahedra. In this case, the substructure 
of this absorption feature tentatively can be attributed to phonon sidebands.
The exciton scenario is strongly supported by the data of K$_{\rm 2}$OsBr$_{\rm 6}$, 
in which both the Mott gap and the excitonic absorption feature are shifted to lower energy 
by about 0.3-0.4\,eV, see Fig.\ \ref{fig:exciton}b).
In contrast, the intra-$t_{2g}$ excitation energies are very similar in 
K$_{\rm 2}$OsBr$_{\rm 6}$ and K$_{\rm 2}$OsCl$_{\rm 6}$, as discussed in Sect.\ \ref{sec:Br}.

The two other below-gap absorption bands without a counterpart in RIXS are the two tiny 
features in $\sigma_1(\omega)$ at about 0.95 and 1.25\,eV, see Fig.\ \ref{fig:IR_elli}.
We attribute also these bands to inter-site excitations. 
They can be explained as a combination and overtone of the intra-$t_{2g}$ excitations 
at about 0.35\,eV and 0.65\,eV, 
i.e., the simultaneous excitation of intra-$t_{2g}$ excitations on two neighboring Os sites. 
In the honeycomb compound $\alpha$-RuCl$_3$ it has been shown that such double or even 
triple excitations may carry sizable spectral weight \cite{Warzanowski20}. Due to 
interaction effects, the peak energies do not have to perfectly match the simple sum 
of the individual excitation energies.

\begin{figure}[t]
	\centering
	\includegraphics[width=\columnwidth]{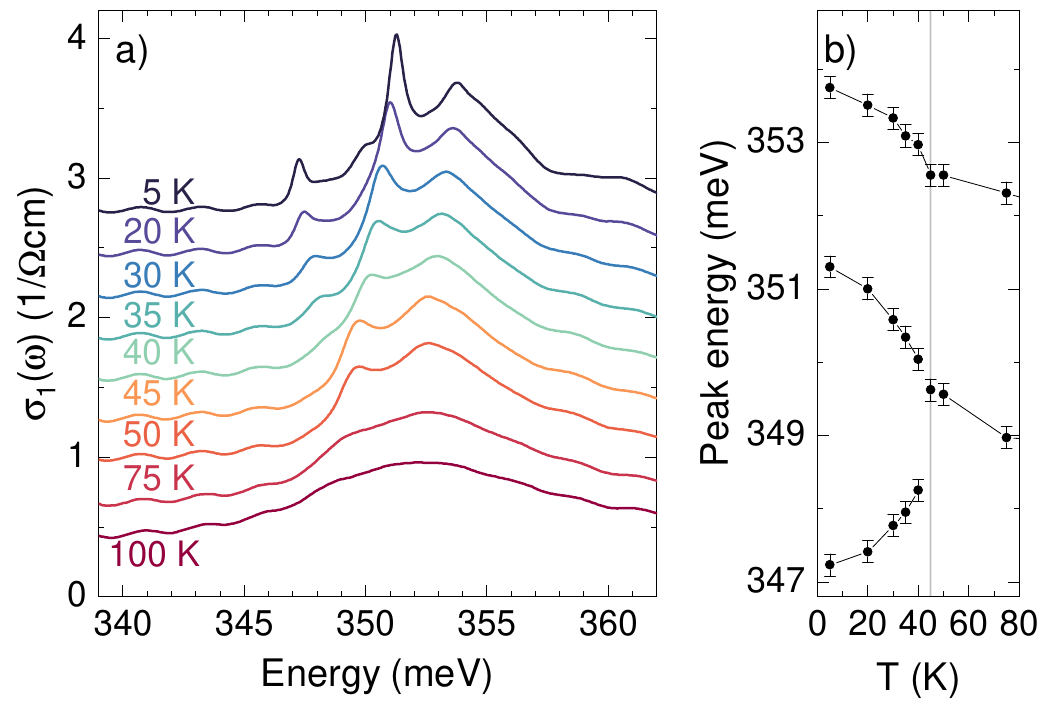}
	\caption{{\bf Temperature dependence of $\sigma_1(\omega)$ across the structural phase transition at 45\,K.} 
	a) Excitations to $J$\,=\,1, cf.\ Fig.\ \ref{fig:sigma_bands}a).
	Data have been offset for clarity.
	b) Corresponding peak energies as a function of temperature. 
	The peak splitting below 45\,K reflects the non-cubic crystal field. 
	}
	\label{fig:non-cubic}
\end{figure}

\subsubsection{Intra-$t_{2g}$ excitations in $\sigma_1(\omega)$}

Considering the intra-$t_{2g}$ excitations, see Figs.\ \ref{fig:t2g_RIXS_IR} and \ref{fig:sigma_bands}, 
the optical data show a rich structure where RIXS finds a single peak. 
Apart from the very different energy resolution, the selection rules and hence the excitation 
mechanisms are  different. 
K$_2$OsCl$_6$ shows inversion symmetry, therefore the electric-dipole matrix element for a 
local, even-parity $d$-$d$ excitation vanishes. Finite spectral weight appears in 
$\sigma_1(\omega)$ based on either a phonon-assisted process, or in the magnetic dipole channel, 
or due to electric quadrupolar or higher-order moments. 
We exemplify this by the data on the $^1\!A_1$ multiplet, see Fig.\ \ref{fig:sigma_bands}d). 

At 5\,K, we find a tiny zero-phonon magnetic-dipole mode at $E_{\,^1\!A_1}$\,=\,2117\,meV.\@ 
This assignment is based on the temperature dependence of the phonon-assisted electric-dipole 
features observed at $E_{\,^1\!A_1} \pm  E_{\rm ph}$. At 5\,K, we only find modes at 
$E_{\,^1\!A_1} \! + \!  E_{\rm ph}$ for different symmetry-breaking phonons with, e.g., 
$E_{\rm ph}$\,=\,7\,meV, 13\,meV, or 21\,meV, as indicated in Fig.\ \ref{fig:sigma_bands}d) 
by the right ends of the horizontal black lines. 
With increasing temperature, these phonon modes become thermally populated and 
the corresponding phonon-annihilation features appear at $E_{\,^1\!A_1}-E_{\rm ph}$, 
strongly supporting our assignment. The spectral weight of the phonon-annihilation peaks 
is governed by the Boltzmann factor. Hence peaks at lower energy 
$E_{\,^1\!A_1}-E_{\rm ph}$ with larger $E_{\rm ph}$ become noticeable at higher temperature. 
Further features above 2150\,meV correspond to a progression of phonon sidebands in a 
vibronic Franck-Condon picture \cite{Henderson,Rueckamp05}. 
Typically, this rich phonon-related pattern of crystal-field excitations can be resolved 
in measurements of transition-metal ions substituted into a host crystal \cite{Kozikowski83}. 
In single crystals, this pattern usually is washed out. 
Exceptions have been observed in quasimolecular crystals such as K$_3$NiO$_2$ 
with isolated NiO$_2$ units \cite{Hitchman}, and the rich optical spectra of K$_2$OsCl$_6$ 
most probably reflect the weak interactions between well separated OsCl$_6$ octahedra.
A detailed assignment of the sidebands in K$_2$OsCl$_6$ has been discussed 
previously \cite{Kozikowski83}.

Similar to the analysis of the $^1\!A_1$ peak, we use the optical data to find the 
bare electronic excitation energies of the intra-$t_{2g}$ excitations at 350, 609, 
and 1311\,meV.\@ The value 350\,meV corresponds to the energy of 
a feature that can be assigned to a magnetic-dipole transition from 
$J$\,=\,0 to 1, i.e., $\Gamma_1$ to $\Gamma_4$ \cite{Homborg82}. 
This peak splits below 45\,K, see Fig.\ \ref{fig:non-cubic} and discussion 
below, but for comparison with theory, we employ the value at 50\,K in the cubic phase. 
The two further energies 609 and 1311\,meV are deduced from 
the temperature dependence of those absorption bands in 
$\sigma_1(\omega)$, in agreement with Ref.\ \cite{Kozikowski83}. 

\begin{figure}[t]
	\centering
	\includegraphics[width=\columnwidth]{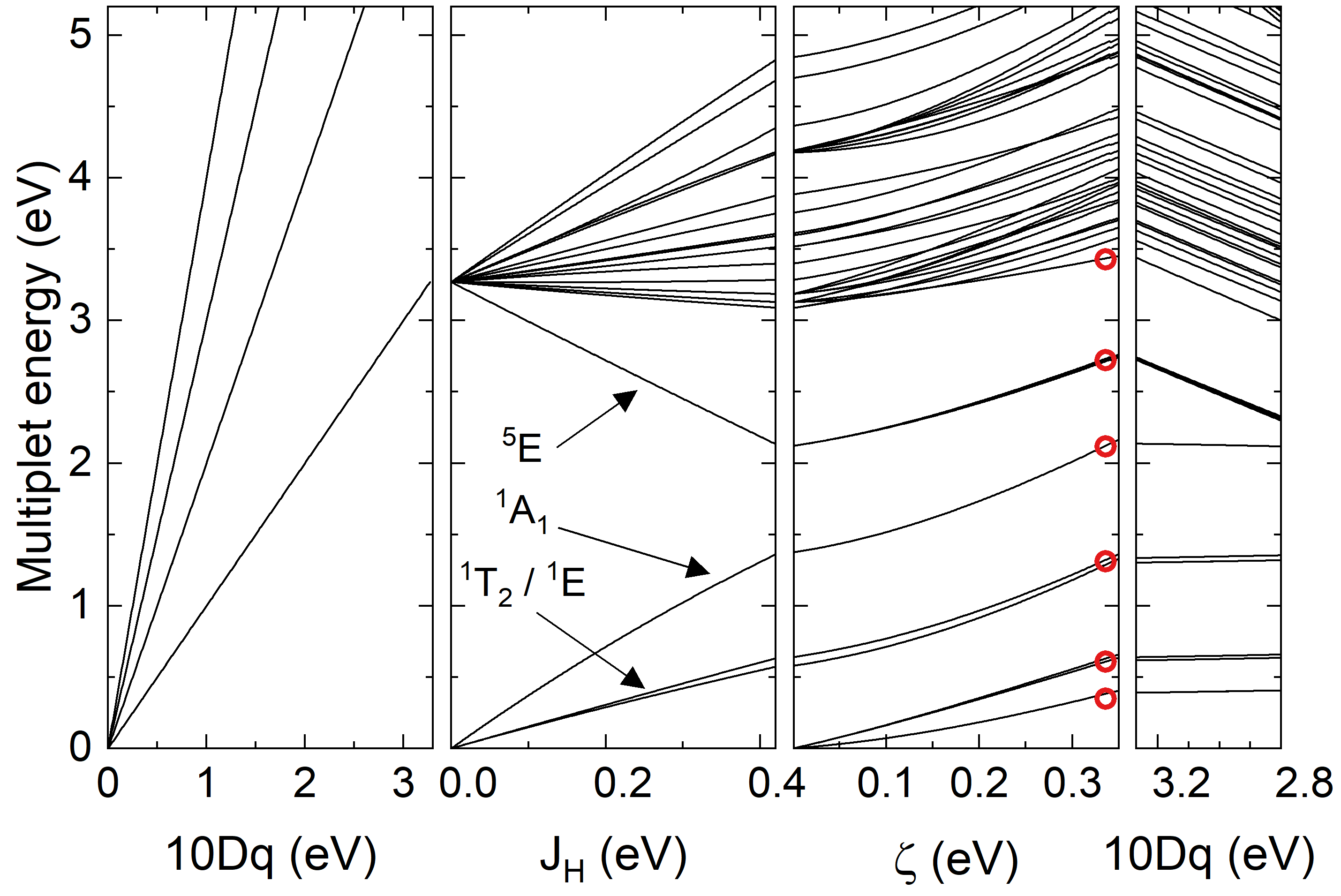	}
	\caption{{\bf Energies of on-site $d$-$d$ excitations for a $5d^4$ configuration 
	in a cubic crystal field.}
	Left: First switching on $10\,Dq$ separates the $t_{2g}^4$ states at zero energy 
	from $t_{2g}^3 e_g^1$ states and states with more than one electron in the 
	$e_g$ orbitals. 
	Middle left: Effect of $J_H$ for $10\,Dq$\,=\,3.3\,eV and $\zeta$\,=\,0. 
	Here, the $^3T_1$ multiplet forms the ground state.
	Middle right: Spin-orbit coupling $\zeta$ yields a $J$\,=\,0 ground state 
	and four groups of intra-$t_{2g}$ excitations. Red circles denote the 
	experimental energies of K$_{\rm 2}$OsCl$_{\rm 6}$ used to determine the 
	electronic parameters.    
	Right: Effect of a reduction of 10\,$Dq$ for finite $J_H$ and $\zeta$, 
	mimicking the trend from K$_{\rm 2}$OsCl$_{\rm 6}$ to K$_{\rm 2}$OsBr$_{\rm 6}$. 	
	}
	\label{fig:energies}
\end{figure}

To compare the excitation energies found in RIXS and optics, we extract the RIXS peak 
energies for the intra-$t_{2g}$ excitations from a fit that uses a series of Voigt profiles. 
We anchor the RIXS energy scale via the optical result of 2117\,meV for the highest 
intra-$t_{2g}$ excitation energy, as mentioned in Sec.\ \ref{sec:spectroscopy}. 
For the other low-energy RIXS peaks, this yields 
341(1), 607(1), 1311(9), and 1373(140)\,meV, where the values in parentheses state 
the error bar of the fit.
Remarkably, the first three values agree with the optical data within less than 
10\,meV, see Fig.\ \ref{fig:sigma_bands}. 
This excellent result once more corroborates the local character of these excitations. 
For the small shoulder around 1.37\,eV, the uncertainty of the peak energy is much larger. 
We hence neglect this value for the further analysis.

The effect of the structural phase transition at 45\,K is addressed in 
Fig.\ \ref{fig:non-cubic}. 
In the antifluorite halides, a non-cubic crystal field splitting $\Delta_{\textrm{CF}}$ 
is driven by tilts and rotary or librational modes of the 
metal-ligand octahedra \cite{Oleary70,Mintz79,Khan21,Bertin22}.
In $\sigma_1(\omega)$, the absorption band at about 0.35\,eV exhibits a peak splitting 
below 45\,K.\@  At 5\,K, we observe peaks at 347 and 351\,meV, i.e., 
$\Delta_{\rm exp}$\,$\approx$\,4\,meV.\@ Upon increasing temperature, the splitting 
gradually decreases and the two peaks merge at 45\,K, see Fig.\ \ref{fig:non-cubic}.
A similar behavior is observed for the $^1T_2$ excitation at 1.3\,eV (not shown). 
This peak splitting $\Delta_{\rm exp}$\,$\approx$\,4\,meV of the $J$\,=\,1 band sets 
the energy scale of $\Delta_{\textrm{CF}}$ of K$_2$OsCl$_6$. For instance for a small 
tetragonal field $|\Delta_{\rm CF}| \ll \zeta$, one finds 
$|\Delta_{\rm CF}| \approx 2 \Delta_{\rm exp}$ \cite{Vergara22}.

\subsection{Calculation of energy levels of K$_{\rm 2}$OsCl$_{\rm 6}$}

\begin{table}[t]
	\begin{tabular}{lcccc}
		\toprule
		{multiplet} & {exp.\ Cl (meV)}& {fit Cl (meV)} & {exp.\ Br (meV)}& {fit Br (meV)}\\
		\midrule
		$\Gamma_4(^3T_1)$ & 350 & 383 & 349 & 388 \\
		$\Gamma_5(^3T_1)$ & 609 & 608 & 604 & 604 \\
		$\Gamma_3(^3T_1)$ & --  & 631 & --  & 632 \\
		$^1T_2$          & 1311 & 1289 & 1285 & 1267 \\
		$^1E$            & --   & 1323 & --  & 1312\\
		$^1\!A_1$        & 2117 & 2123 & --  & 2050 \\
		$^5E$            & 2721 & 2716 & 2391& 2381  \\
		$^3E$            & 3428 & 3432 & 3033 & 3042  \\
		\bottomrule
	\end{tabular}
	\caption{{\bf Experimental and calculated excitation energies.} 
		The four intra-$t_{2g}$ energies are taken from the optical data, the two $e_g$ 
		levels above 2.7\,eV from RIXS.\@ 
		Dashes denote peaks that are not resolved in the experiment. 
		For K$_{\rm 2}$OsCl$_{\rm 6}$ (K$_{\rm 2}$OsBr$_{\rm 6}$), the parameters of the 
		$d$-shell model are 10$Dq$\,=\,3.27\,eV (2.85\,eV), $\zeta$\,=\,0.336\,eV (0.325\,eV), 
		$F^2$\,=\,3.73\,eV (3.90\,eV), and $F^4$\,=\,2.22\,eV (2.01\,eV), 
		which corresponds to $J_H$\,=\,0.425\,eV (0.422\,eV).
\label{tab:calc}
}
\end{table}

\begin{figure}[t]
	\centering
	\includegraphics[width=\columnwidth]{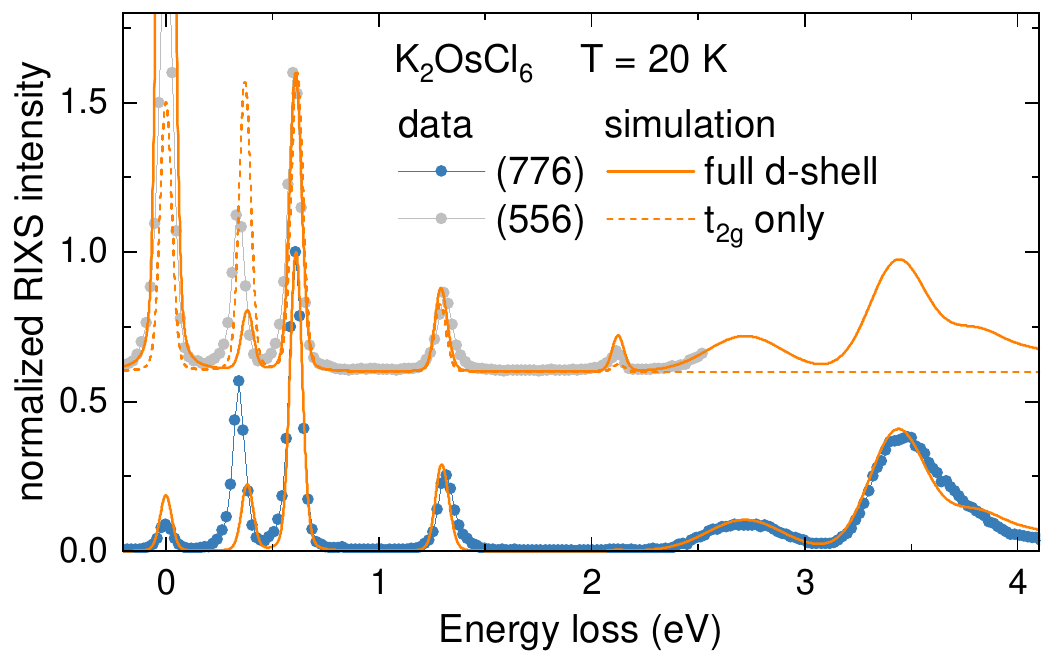}
	\caption{{\bf Measured and calculated RIXS spectra of K$_2$OsCl$_6$.} 
		Solid orange lines: Simulation based on the fit result for the 
		full $d$-shell model, cf.\ Tab.\ \ref{tab:calc}. 
		To account for the polarization dependence of the $^1A_1$ transition 
		at 2.1\,eV, we consider two different values of $\mathbf{q}$.    		
		Dashed: Corresponding result of the Kanamori model for 
		$J_H^{\rm eff}$\,=\,0.28\,eV and $\zeta^{\rm eff}$\,=\,0.41\,eV.\@ 
		Peak widths are adapted to the data. Offsets have been used for clarity.
	}
	\label{fig:simulation}
\end{figure}

For the analysis of the electronic parameters, we stick to cubic symmetry. In a single-site 
model, the relevant parameters for the energy levels are the cubic crystal-field splitting 
10\,$Dq$ between $t_{2g}$ and $e_g$ levels, spin-orbit coupling $\zeta$, and the 
Coulomb interaction within the $5d$ shell. 
In spherical approximation, the latter can be described by the Slater integrals 
$F^2$ and $F^4$, which are used to calculate Hund's coupling 
$J_H$\,=\,1/14\,$\left(F^2 \! + \! F^4\right)$ within the entire $5d$ shell \cite{Georges13}. 
Note that $F^0$, equivalent to Hubbard $U$, does not contribute to the energy splitting 
for a single site with a fixed number of electrons.  
We calculate the multiplet energies using \textit{Quanty} \cite{Haverkort12,Haverkort16}. 
Figure\ \ref{fig:energies} displays the behavior as a function of 10\,$Dq$, $J_H$, and 
$\zeta$. Starting with $J_H$\,=\,$\zeta$\,=\,0, the cubic crystal field 
yields a $t_{2g}^4$ ground state and raises the excitation energies of the 
$t_{2g}^{4-n}\,e_g^n$ states with $n$\,=\,1 to 4 electrons in the $e_g$ subshell. 
Switching on $J_H$ splits each of these five branches. We focus on the 15 $t^4_{2g}$ states. 
These are split into the $^3T_1$ ground state and the $^1T_2$, $^1E$, and $^1A_1$ excited 
states. 
The excitation energies of roughly 3/2\,$J_H$ and 4\,$J_H$ are approximately 20-25\,\% 
lower than for 10\,$Dq$\,=\,$\infty$, reflecting a small but finite admixture of 
$e_g$ character. 
Finally, spin-orbit coupling $\zeta$ causes a further fanning out of the energies 
and splits the $^3T_1$ ground state into three branches. In the limit of 
10\,$Dq$\,=\,$J_H$\,=\,$\infty$ these correspond to the eigenstates $J$\,=\,0, 1, and 2.
The intra-$t_{2g}$ excitations within the $t_{2g}^4$ states form four groups of excitations 
that cover the range up to about 2\,eV.\@ 
The lowest $t_{2g}^3 e_g^1$ level corresponds to the high-spin $^5\!E$ state with $S$\,=\,2. 
With all four spins being parallel, it is strongly favored by $J_H$. 
Even though the $^5\!E$ state is well separated from other levels, cf.\ Fig.\ \ref{fig:energies}, 
the width of this $t_{2g}$-to-$e_g$ excitation is much larger than observed for the intra-$t_{2g}$ 
peaks, see Fig.\ \ref{fig:simulation}. This suggests a finite mixing with excitations across 
the Mott gap at 2.2\,eV.

Optical studies of Os$^{4+}$ impurities in different host crystals have reported different sets 
of electronic parameters \cite{Dorain68, Rahman71, Kozikowski80,Piepho72}.
Also for optical data on single crystals of K$_2$OsCl$_6$, it has been found that 
the determination of the electronic parameters is difficult \cite{Kozikowski83}, 
foremost because the crystal-field splitting 10\,$Dq$ is hard to obtain 
from the optical data. 
The combined approach of RIXS and optics is pivotal here to resolve these ambiguities. 

\begin{figure}[t]
	\centering
	\includegraphics[width=\columnwidth]{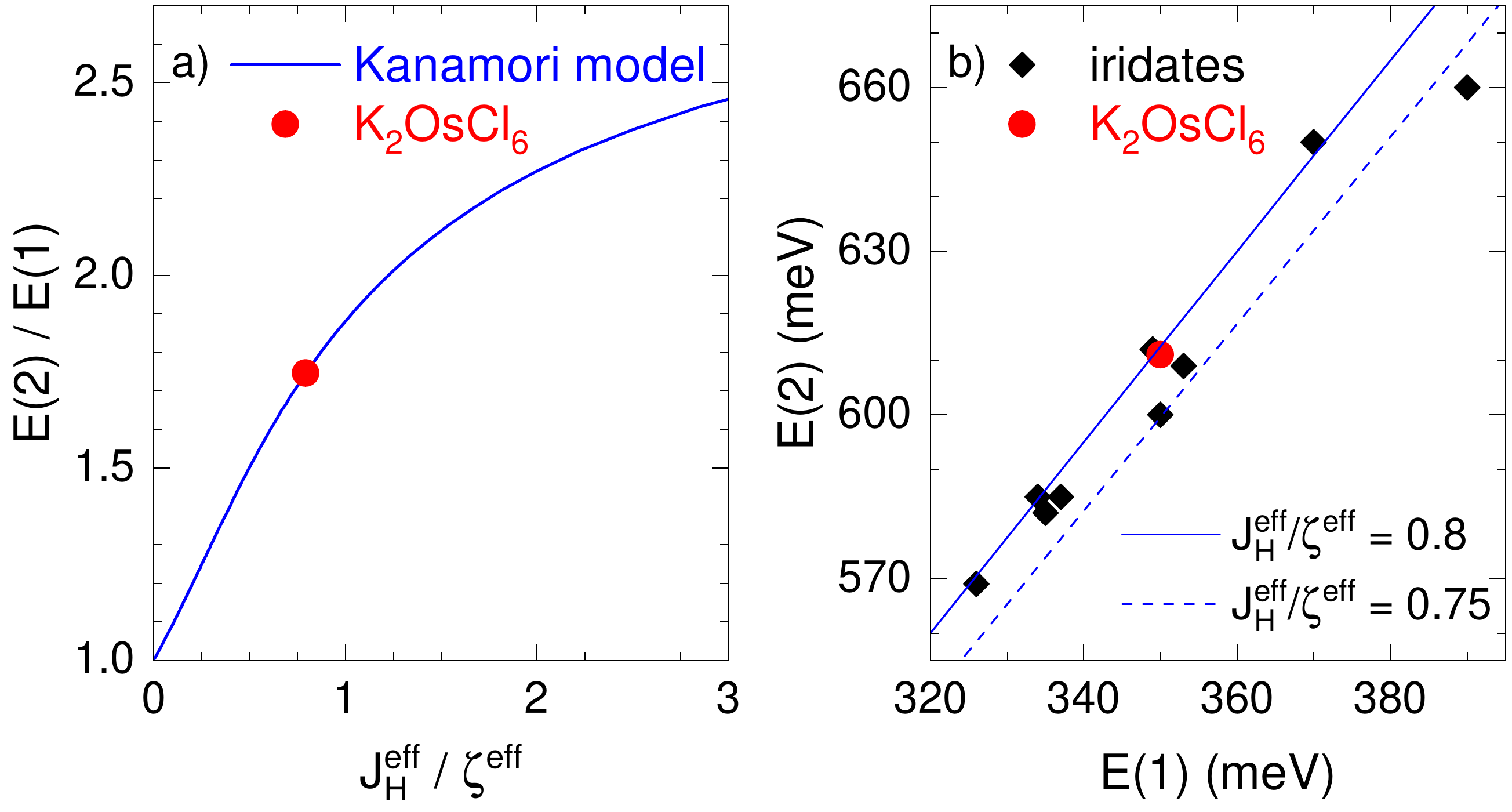}
	\caption{{\bf Analysis based on the $t_{2g}$-only Kanamori model.} 
		a) The ratio $E(2)/E(1)$ varies between 1 for 
		$J_H^{\rm eff}/\zeta^{\rm eff} \! \rightarrow \! 0$ and 3 for 
		$J_H^{\rm eff}/\zeta^{\rm eff} \! \rightarrow \! \infty$. 
		For K$_{\rm 2}$OsCl$_{\rm 6}$, it yields $J_H^{\rm eff}/\zeta^{\rm eff}$\,$\approx$\,3/4. 
		b) Comparison of the two lowest excitation energies in 
		K$_{\rm 2}$OsCl$_{\rm 6}$ and related $5d^4$ iridates \cite{Nag18,Aczel22,Kusch18,Yuan17}. For all of them, 
		$J_H^{\rm eff}/\zeta^{\rm eff}$ is very similar. 
	}
	\label{fig:Kanamori}
\end{figure}

To determine the parameters of the $d$-shell model for K$_2$OsCl$_6$, we consider the 
four energies of the intra-$t_{2g}$ excitations from the optical data and the 
$t^3_{2g}\,e_g^1$ excitations observed at 2.72 and 3.43\,eV in RIXS, 
see red circles in Fig.\ \ref{fig:energies}.
We neglect peaks at higher energies where an unambiguous assignment is hindered by the 
large number of similar excitation energies. 
In the fit, we minimize the absolute difference between experimental peak energies and 
calculated ones. This yields the parameters 
$10\,Dq$\,=\,3.3\,eV, $\zeta$\,=\,0.34\,eV, $F^2$\,=\,3.7\,eV, and $F^4$\,=\,2.2\,eV, 
resulting in $J_H$\,=\,0.43\,eV.\@ 
Note that the value of $10\,Dq$ agrees very well with our simple estimate discussed above. 
Figure \ref{fig:simulation} compares the calculated result for direct $L_3$-edge 
RIXS \cite{Ament11} for this parameter set with the experimental data. 
Overall, the peak energies as well as the relative peak intensities are well described.

The calculated excitation energies are given in Tab.\ \ref{tab:calc}. 
For the splittings of the excitations at about 0.6 and 1.3\,eV,
the model predicts values of 20-35\,meV in cubic symmetry, which is well below 
the RIXS resolution limit and agrees with the energy scale observed in the optical data. 
Concerning the fitted energies, the deviations are less than 7\,meV for four of the 
six energies, an excellent result. 
Upon a closer look, the largest deviation of 33\,meV is found for the peak lowest 
in energy, and the intensity of this feature is underestimated.
A similar observation has been reported for $4d^4$ K$_2$RuCl$_6$, where it has been 
discussed in terms of a possible role of vibronic coupling, i.e., a dynamic Jahn-Teller 
splitting of the excited $J$\,=\,1 triplet state \cite{Takahashi21,Iwahara23}. 
For K$_{\rm 2}$OsCl$_{\rm 6}$, our optical data resolve the vibronic sidebands in the excited 
states,  cf.\ Fig.\ \ref{fig:sigma_bands}, but do not hint at a particular importance of 
vibronic effects for the feature around 0.35\,eV.\@ 
We find, however, that a small change in the parameters can eliminate 
this apparent shortcoming of the model. In a second fit, we minimize the \textit{relative} 
energy difference between experiment and model. This results in the very similar 
parameter set $10\,Dq$\,=\,3.4\,eV, $\zeta$\,=\,0.34\,eV, and $J_H$\,=\,0.42\,eV and 
yields the energies 358, 600, 1278, and 2139\,meV for the four lowest excitations. 
For all four, the relative difference is less than 2.6\,\%, a very reasonable result. 
In terms of absolute energy differences, the deviation between fit and experiment is reduced 
to 8\,meV for the peak at 350\,meV while the description of the other peaks is slightly worse 
compared to the first fit.
Finally, we have checked that inclusion of the Cl 3$p$ orbitals and ligand-to-metal 
charge-transfer processes \cite{Haverkort12} yields very similar results.  
Overall, we find $J_H/\zeta$\,$\approx$\,1.3 which puts K$_2$OsCl$_6$ in the 
intermediate regime between $LS$ coupling with $J_H/\zeta \! \rightarrow \! \infty$ and 
$jj$ coupling for $J_H/\zeta \! \rightarrow \! 0$.

\begin{figure}[t]
	\centering
	\includegraphics[width=\columnwidth]{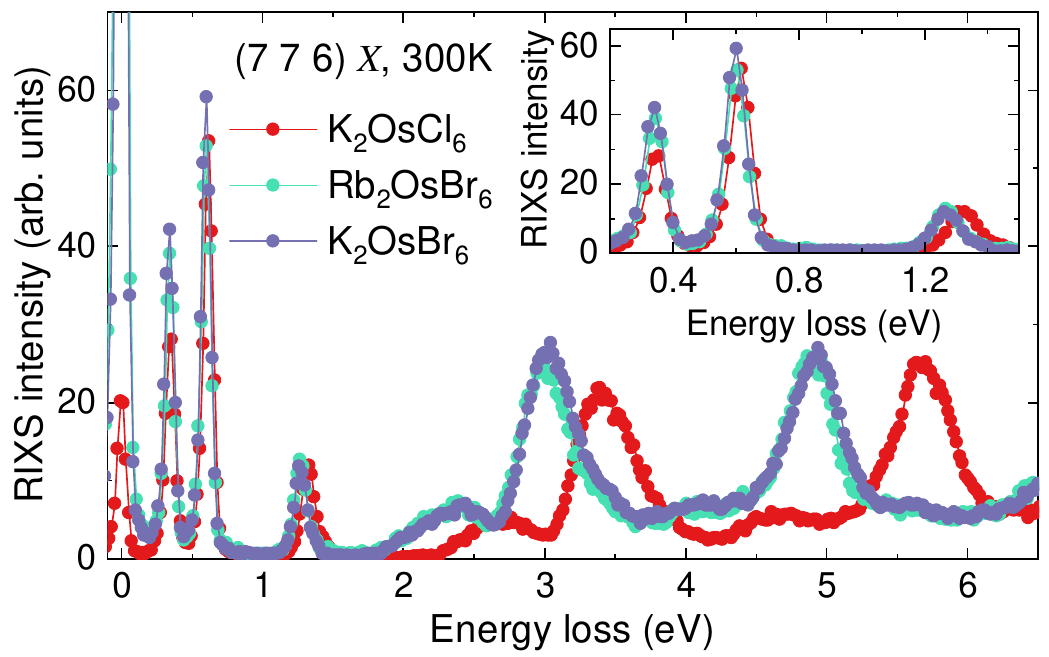}
	\caption{{\bf Comparison of RIXS spectra of K$_2$OsCl$_6$, K$_2$OsBr$_6$, and Rb$_2$OsBr$_6$.} 
	The difference in size and electronegativity between Br and Cl ions alters the 
	crystal-field and charge-transfer excitations. In comparison, the intra-$t_{2g}$ peaks 
	are less affected by the change of the ligand, as highlighted in the inset.   	
	}
	\label{fig:bromide}
\end{figure}

\subsection{Parameters in the Kanamori model}

Thus far we discussed a model that takes the entire $d$ shell into account. 
In contrast, the related $5d^4$ iridate data in Refs.\ \cite{Yuan17,Nag18,Kusch18,Aczel22} 
were analyzed using the Kanamori model that assumes 10\,$Dq$\,=\,$\infty$, i.e., 
it considers only $t_{2g}$ orbitals. For a comparison, it is important to quantify 
how the parameter values depend on the chosen model. 
In the Kanamori model, the only parameters are $\zeta^{\rm eff}$ and 
$J_H^{\rm eff}$\,=\,$3/49\, F^2  +  20/441 \, F^4$ \cite{Georges13}. 
With $10\,Dq$\,=\,$\infty$, the Kanamori model restores the degeneracy of $^1T_2$ and $^1E$ 
around 1.3\,eV and of the $\Gamma_3$ and $\Gamma_5$ states of $J$\,=\,2 around 0.6\,eV, 
as mentioned in Sec.\ \ref{sec:intrat2g}. Hence the model has only four intra-$t_{2g}$ 
excitation energies. For simplicity, we call these energies $E(1)$ to $E(4)$, 
in ascending order. Simple expressions for $E(1)$ to $E(4)$ are given in 
Ref.\ \cite{Iwahara23}.

A fit of the four intra-$t_{2g}$ excitation energies with the $t_{2g}$-only 
Kanamori model yields $J_H^{\rm eff}$\,=\,0.28\,eV and $\zeta^{\rm eff}$\,=\,0.41\,eV
and the energies 371, 616, 1288, and 2125\,meV.\@ A corresponding simulated RIXS spectrum 
is shown in Fig.\ \ref{fig:simulation}. Similar to the result obtained for the 
full $d$ shell, the relative deviation between fit and data is largest for the lowest mode. 
Again, a slightly different set of parameters yields excellent agreement for the 
lowest excitation energies. 
Choosing $J_H^{\rm eff}$\,=\,310\,meV and $\zeta^{\rm eff}$\,=\,399\,meV describes the 
lowest three peaks of K$_{\rm 2}$OsCl$_{\rm 6}$ within 1\,meV and $E(4)$ 
within 6\,\%. Overall, this suggests $J_H^{\rm eff}/\zeta^{\rm eff}$\,$\approx$\,3/4.

Another possibility to determine this quantity is given by the ratio $E(2)/E(1)$, 
which varies from 3 for $LS$ coupling to 1 for $jj$ coupling, see Fig.\ \ref{fig:Kanamori}a).
For K$_{\rm 2}$OsCl$_{\rm 6}$ with 609\,meV/350\,meV\,$\approx$\,1.7, this corroborates 
$J_H^{\rm eff}/\zeta^{\rm eff}$\,$\approx$\,3/4. 
Remarkably, very similar values of the ratio $E(2)/E(1)$ have been reported for several 
$5d^4$ iridates, see Fig.\ \ref{fig:Kanamori}b), highlighting the close 
relationship of the properties of this series of $5d^4$ compounds.

As long as the corresponding model is specified, both $J_H/\zeta$\,$\approx$\,1.3 for the entire $5d$ shell and 
$J_H^{\rm eff}/\zeta^{\rm eff}$\,$\approx$\,3/4 in the $t_{2g}$-only model 
are valid. For $J_H$\,$\approx$\,$\zeta$, the two models predict 
\begin{eqnarray}
E(J=2) & \approx & \frac{3}{2}\zeta \, \left( 1 + \frac{2\zeta}{10\,Dq} \right) 
\\
E(2) & \approx & \frac{3}{2}\zeta^{\rm eff}
\end{eqnarray}
for the second excitation energy and hence $\zeta^{\rm eff}/\zeta$\,$\approx$\,1.2 
in the $5d^4$ compounds. Additionally, the different definitions of Hund's coupling 
yield $J_H^{\rm eff}/J_H$\,$\approx$\,0.77. 
Concerning the energies of the intra-$t_{²g}$ excitations, the 
$t_{2g}$-only Kanamori model and the model for the full $d$ shell work equally well. 
The main shortcoming of both models is the intensity of the lowest excitation, 
which is too small in the $d$-shell model but too large in the Kanamori model, 
see Fig.\ \ref{fig:simulation}. This suggests that the intensity of the lowest RIXS peak 
depends sensitively on 10\,$Dq$ and on the admixture of $e_g$ character.

\section{Results on the bromides}
\label{sec:Br}

\begin{figure}[t]
	\centering
	\includegraphics[width=\columnwidth]{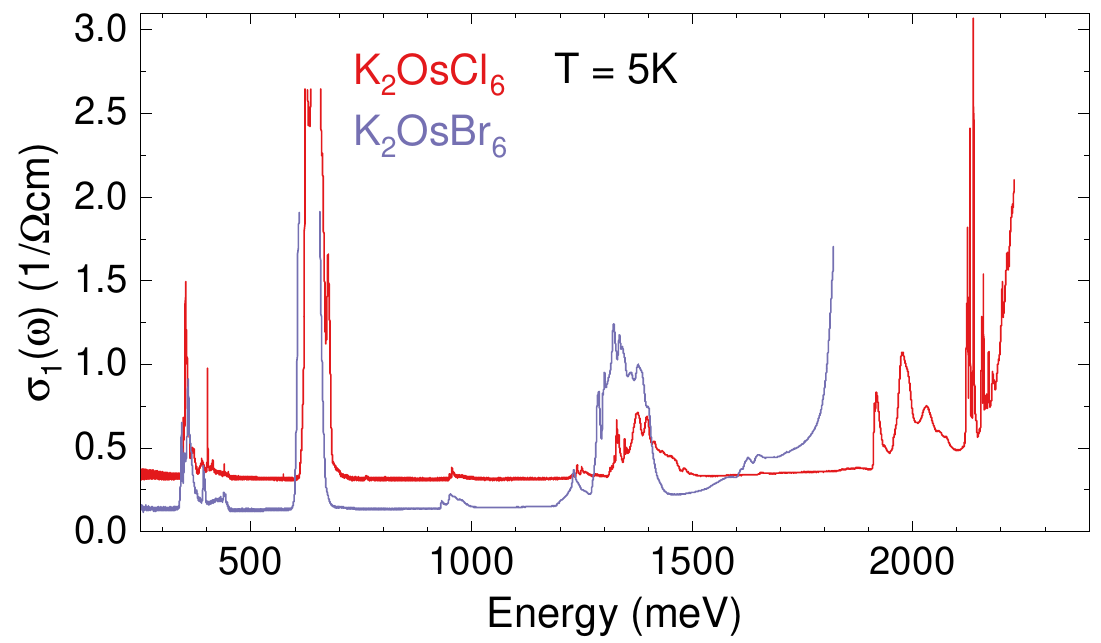}
	\caption{{\bf Optical conductivity of K$_2$OsCl$_6$ and K$_2$OsBr$_6$.} 
	The main change is the shift of the Mott gap from 2.2\,eV to 1.8\,eV.\@ 
	The small offset in the K$_{\rm 2}$OsCl$_{\rm 6}$ data indicates a possible surface issue. 
	Note that we polished this sample with CeO$_2$ in propanol, 
	while the K$_{\rm 2}$OsBr$_{\rm 6}$ crystal was measured as grown.
	}
	\label{fig:bromide_sig}
\end{figure}

RIXS spectra of K$_2$OsCl$_6$, K$_2$OsBr$_6$, and Rb$_2$OsBr$_6$ are compared in 
Fig.\,\ref{fig:bromide}. We find a close resemblance of the RIXS data of K$_2$OsBr$_6$ 
and Rb$_2$OsBr$_6$, i.e., a very small effect of the $A$ ion. Hence we focus 
on the comparison of K$_2$OsBr$_6$ and K$_{\rm 2}$OsCl$_{\rm 6}$. 
The behavior is different above and below 2\,eV.\@ 
In RIXS we find the intra-$t_{2g}$ excitation energies of K$_{\rm 2}$OsBr$_{\rm 6}$ 
at 0.34, 0.60, and 1.27\,eV.\@ Compared to K$_{\rm 2}$OsCl$_{\rm 6}$, these are reduced 
by about 1\,\%, 2\,\%, and 3\,\%, respectively, i.e., they are almost unaffected 
by the choice of the halide. In contrast, the peak energies associated to the $e_g$ 
and charge-transfer excitations are reduced by about 12-14\,\% in K$_{\rm 2}$OsBr$_{\rm 6}$ 
with respect to the chloride. 
More precisely, the peak energies are 2.4 and 3.0\,eV for the strongest $t^3_{2g}e_g^1$ 
transitions and 4.0 and 4.9\,eV for the charge-transfer excitations 
$|5d_{\rm Os}^4 \, 4p_{\rm Br}^6 \rangle$\,$\rightarrow$\,$|5d_{\rm Os}^5 4p_{\rm Br}^5 \rangle$. 
The very similar intensity profile of the two compounds indicates that the main effect 
at high energy is captured by a renormalization of the energy scale, i.e., of 
10\,$Dq$ and $\Delta_{\mathrm{CT}}$. 
Both parameters are affected by the difference in ionic size and electronegativity. 

In analogy to the discussion of K$_{\rm 2}$OsCl$_{\rm 6}$ above, we identify 
$\Delta_{\mathrm{CT}}^{\rm Br}$\,=\,4.0\,eV via the lowest charge-transfer excitation. 
Furthermore, a fit of the excitation energies using multiplet calculations yield 
the parameters $10Dq^{\rm Br}$\,=\,2.9\,eV, $\zeta^{\rm Br}$\,=\,0.33\,eV, and 
$J_{H}^{\rm Br}$\,=\,0.42\,eV.\@ 
Compared to K$_{\rm 2}$OsCl$_{\rm 6}$, the spin-orbit coupling constant is reduced by about 4\,\%, 
while the change of $J_H$ is negligible, see Tab.\ \ref{tab:calc}.
The effect of a reduction of 10\,$Dq$ on the energy levels of the multiplet model 
is shown in the right panel of Fig.\ \ref{fig:energies}. While the $t_{2g}^3 e_g^1$ 
levels decrease linearly in energy, the $t_{2g}^4$ states hardly change.
In general, lower values of 10\,$Dq$ and $\Delta_{\mathrm{CT}}$ indicate a stronger 
admixture of $e_g$ and $4p$ ligand character into the $t_{2g}$ states. 
This, in turn, reduces the effective value of $\zeta$.  
However, the large ratio of 10\,$Dq/\zeta \approx 10$ explains why the sizable 
reduction of 10\,$Dq$ has only a small effect on $\zeta$.

The optical conductivity $\sigma_1(\omega)$ of K$_2$OsBr$_6$ is depicted in 
Fig.\ \ref{fig:bromide_sig} and in Fig.\ \ref{fig:exciton}b). Note that the larger thickness 
$d^{\rm Br}$\,=\,170\,$\mu$m of the bromide sample limits the accessible values of 
$\sigma_1(\omega)$ to below 2\,$(\Omega \mathrm{cm})^{-1}$, somewhat lower than in the 
thinner chloride sample. The most pronounced difference to the data of K$_{\rm 2}$OsCl$_{\rm 6}$ 
is the value of the Mott gap, which is shifted down by about 0.4\,eV to 1.8\,eV at 5\,K.\@
This shift masks the $^1A_1$ excitation which occurs at 2.1\,eV in K$_{\rm 2}$OsCl$_{\rm 6}$.  
Along with the Mott gap, also the excitation energy of the exciton is reduced, 
as already discussed in connection with Fig.\ \ref{fig:exciton}. 
At lower energy, the small shifts of the intra-$t_{2g}$ excitations of less than 
1\,\% for the lowest excitation and about 1-2\,\% for the bands at 0.6 and 1.3\,eV 
agree with the RIXS results.

\section{Conclusion}

In conclusion, we have probed the local electronic structure of the 5$d^4$ 
hexahalogenoosmates K$_2$OsCl$_6$, K$_2$OsBr$_6$, and Rb$_2$OsBr$_6$ 
with magnetic and spectroscopic methods. These measurements reveal non-magnetic 
$J$\,=\,0 behavior, as expected for a clean 5$d^4$ system in which both defects 
and exchange interactions are negligible. 
Combining RIXS and optical spectroscopy, we can assign the multiplet excitation 
energies with high accuracy and extract the electronic parameters by comparison 
with local multiplet calculations. RIXS at the transition-metal $L$ edge is most 
sensitive to on-site $d$-$d$ excitations, which in the optical data give rise to weak, 
typically phonon-assisted features. For the antifluorite-type Os halides, a central 
advantage of RIXS is the ability to determine 10\,$Dq$ via the observation of 
local excitations into the $e_g$ subshell also above the Mott gap. Due to the 
weak interactions between Os$X_6$ ($X$\,=\,Cl, Br) octahedra, both the $e_g$ excitations 
and the charge-transfer excitations are narrow and very well defined in RIXS.\@
As a complementary technique, optical spectroscopy allows us to determine the 
small non-cubic crystal-field splitting of 4\,meV below 45\,K in K$_{\rm 2}$OsCl$_{\rm 6}$. 
Furthermore, it is sensitive to intersite processes such as excitations across 
the Mott gap at 2.2\,eV and reveals an exciton around 1.9-2.1\,eV.\@ 
These results establish the presented compounds as realizations of clean cubic 
$J$\,=\,0 systems in the intermediate coupling regime with $J_H/\zeta$\,$\approx$\,1.3 
or $J_H^{\rm eff}/\zeta^{\rm eff}$\,$\approx$\,3/4.
Accordingly, Coulomb interaction and spin-orbit coupling have to be taken on equal 
footing. This value of $J_H^{\rm eff}/\zeta^{\rm eff}$ is very similar to results 
reported for $5d^4$ iridates \cite{Nag18,Aczel22,Kusch18,Yuan17},
highlighting the close relationship between these compounds. 
The comparison of the data from chloride and bromide samples shows a 20\,\% decrease 
of the Mott gap as well as a reduction of about 12-14\,\% of 10\,$Dq$ and of the charge-transfer 
energy $\Delta_{\rm CT}$. Due to the large size of 10\,$Dq$ and $\Delta_{\mathrm{CT}}$ 
with respect to $J_H$ and $\zeta$, this sizable change has only a marginal effect on 
intra-$t_{2g}$ energies, which are reduced by 1-3\,\%. 
In contrast to previous studies on $5d^4$ $J$\,=\,0 compounds, where the 
determination of $\zeta$ and $J_H$ has turned out to be difficult, we find 
that both chloride and bromide samples are well described by $\zeta$\,=\,0.33-0.34\,eV  
and $J_H$\,=\,0.42-0.43\,eV.\@ These values may serve as a solid reference for future studies 
on Os compounds. 
\medskip

\begin{acknowledgments}
We gratefully acknowledge H. Schwab for experimental assistance in sample characterization 
and the European Synchrotron Radiation Facility for providing beam time and technical support. 
Furthermore, we acknowledge funding from the Deutsche Forschungsgemeinschaft 
(DFG, German Research Foundation) through Project No.\ 277146847 -- CRC 1238 
(projects A02, B01, B02, B03). 
\end{acknowledgments}

\end{document}